\documentclass[11pt]{article}

\usepackage[utf8x]{inputenc}
\usepackage{amssymb}
\usepackage{amsmath}
\usepackage{amsthm}
\usepackage{graphicx}
\usepackage{epsfig}
\usepackage{float}
\usepackage{fancyhdr}
\usepackage{enumerate}
\usepackage{hyperref}
\usepackage{times}
\usepackage{caption}
\usepackage{subfig}
\usepackage{tikz}
\usepackage{siunitx,booktabs,array}

\newtheorem{Definition}{Definition}[part]
\newtheorem{Proposition}{Proposition}[part]
\newtheorem{Assumption}{Assumption}[part]

\newtheorem{Remark}{Remark}[part]

\newtheorem{Notation}{Notation}[part]

\newcolumntype{P}[1]{>{\centering\arraybackslash}p{#1}}
\newcolumntype{M}[1]{>{\centering\arraybackslash}m{#1}}

\def \N{\mathbb{N}}
\def \R{\mathbb{R}}

\def\1{\mathds{1}}

\def\beqs{\begin{eqnarray*}}
\def\enqs{\end{eqnarray*}}
\def\beq{\begin{eqnarray}}
\def\enq{\end{eqnarray}}

\newcommand{\nc}{\newcommand}
\nc{\esssup}{\mathop{\mathrm{ess\,sup}}}

\definecolor{Gcolor}{rgb}{1,0,0.5}

\begin{document}
\title{A Self-Exciting Modelling Framework for Forward Prices in Power Markets}
\author{Giorgia Callegaro\thanks{Universit\'a di Padova, Dipartimento di Matematica Applicata. Email: gcallega@math.unipd.it} \,\,\,\quad  Andrea Mazzoran\thanks{Universit\'a di Padova, Dipartimento di Matematica Applicata. Email: mazzoran@math.unipd.it} 
 \,\,\,\quad Carlo Sgarra\thanks{Politecnico di Milano, Dipartimento di Matematica. CORRESPONDING AUTHOR. Email: carlo.sgarra@polimi.it}}

\date{\today}

\maketitle
\begin{abstract}
We propose and investigate two model classes for forward power price dynamics, based on continuous branching processes with immigration, and on Hawkes processes with exponential kernel, respectively. The models proposed exhibit jumps clustering features. Models of this kind have been already proposed for the spot price dynamics, but the main purpose of the present work is to investigate the performances of such models in describing the forward dynamics. We adopt a Heath-Jarrow-Morton approach in order to capture the whole forward curve evolution. By examining daily data in the French power market, we perform a goodness-of-fit test and we present our conclusions about the adequacy of these models in describing the forward prices evolution.
\end{abstract}

\vspace{1cm}


%
%

\section{Introduction}
Energy markets, and in particular, electricity markets, exhibit very peculiar features. The historical series of both futures and spot prices include seasonality, mean-reversion, spikes and small fluctuations. One can alternatively describe the power price dynamics by modelling the spot or the forward price. In the former case, the spot price can be obtained as a limit of the forward price when the maturity is close to the current time, in the latter case it is possible to derive the forward price from the spot by computing the conditional expectation with respect to a suitable risk-neutral measure of the spot price at the maturity.

After the pioneering paper by Schwartz \cite{Schwarz}, where an Ornstein-Uhlenbeck dynamics is assumed to describe the spot price behaviour, several different approaches have been investigated in order to describe the power price evolution. A comprehensive literature review until $2008$ is offered in the book by Benth et al. \cite{BBK}. A similar effort has been devoted to identify reliable models for the forward price dynamics, and a huge amount of literature is available focusing on Heath-Jarrow-Morton type models as in Benth et al. \cite{BPV18} and in Filimonov et al. \cite{LPV18}, in the attempt to provide a description of the whole forward curves dynamics, in analogy with forward interest rates in fixed-income markets as in Heath et al. \cite{HJM}. Some of the classical models proposed include jumps and/or stochastic volatility. Benth and Paraschiv \cite{BP17} propose a random field approach based on Gaussian random fields by adopting the Musiela parametrization in order to describe the forward curve dynamics. Empirical evidence suggests that in many assets prices often jumps appear in cluster, thus requiring the introduction of jump processes exhibiting a clustering or self-exciting behaviour.

Kiesel and Paraschiv \cite{Kiesel} recently presented a systematic empirical investigation of electricity intraday prices and suggested an approach based on Hawkes processes in order to describe the power price dynamics with jump clustering features. Self-exciting features in electricity prices attracted already some attention by several authors: Herrera and Gonzalez \cite{HerGon2014} proposed a self-excited model for electricity spot prices, while Christensen et al. \cite{CHL2009}, Clements et al. \cite{CFH2013} pointed out that time between spikes has a significant impact on the likelihood of future occurrences, thus providing a strong support to models including self-exciting properties.

The large class of models available in the literature describing the power price dynamics is then widening in order to include models exhibiting self-exciting features.

We also mention the paper by Jiao et al. \cite{JMSS}, where a model based on continuous branching processes with immigration for power spot prices was proposed, and the forward prices computed with respect to a suitable structure preserving equivalent martingale measure.

Eyjolfsson and Tj{\o}shteim \cite{Eyjolfsson2018} describe a class of Hawkes processes and present an empirical investigation based on data from UK power market supporting Hawkes-type models for spot prices.

The purpose of the present paper is to investigate if self-exciting features can arise in the power forward prices evolution as well, and in order to perform this investigation we shall focus on two different model classes: the Continuous Branching Processes with Immigration (CBI henceforth) and the Hawkes processes. While CBI processes are always affine, Hawkes processes in general are not, but when the kernel describing the intensity dynamics is of exponential type they are, and this feature makes the Hawkes processes with exponential kernel appealing from the modeling point of view.
By considering the two model classes mentioned before, i.e. CBI and Hawkes processes, we then want to provide the description of the full term structure of power forward prices, following a Heath-Jarrow-Morton approach.

Power is a flow commodity, this meaning that instantaneous forward contracts are not directly traded on the market, but futures (sometimes called flow forward) are. So, in order to perform any kind of inference on the model proposed, it is necessary to extract the relevant information on the forward dynamics included in the futures prices. This can be done by applying suitable optimization procedures proposed in the literature, eventually modified in order to provide the best performances in the case under examination. These procedures are far from trivial from the computational point of view and require a careful implementation of the optimization step. We deliberately chose to work on daily data, in order to show how self-exciting effects can arise not only on a small time scale, but also at a coarser level.

The paper is organized as follows:  in Section \ref{Modeling Framework} we introduce the processes on which our models are based and in Section \ref{Forward Model} we present and discuss the models proposed for the forward power price dynamics.
In Section \ref{Futures-Dynamics} we discuss the dynamics of Futures contracts when the forward dynamics is assumed to be given by the models introduced.
From Sections \ref{Section: Data Analysis} to \ref{Section: Testing the Models}, we provide the theoretical background and numerical results relative to the calibration/parameters' estimation for the model proposed.
In the final section we provide some concluding remarks and discuss future extensions of the present work.

\section{The Modeling Framework}  \label{Modeling Framework}

\subsection{Continuous Branching Processes with Immigration}  \label{CBI}

We now introduce our modeling framework for the electricity price, which is based on stochastic differential equations driven by L\'evy random fields. We consider a L\'evy random field, which is a combination of a Gaussian random measure $W$ and a compensated Poisson random measure $N$ independent of $W$. For a background on such general stochastic equations with jumps, we refer the readers e.g. to Dawson and Li \cite{DawsonLi},  Li and Ma  \cite{LM13} and Walsh \cite{Walsh1980}.

Let us now briefly introduce all the relevant ingredients of our work and recall some preliminary results. We fix a probability space $(\Omega,\mathcal A,\mathbb P)$.
A {\it  white noise} $W$ on $\mathbb R_+^2$ is a Gaussian random measure such that, for any Borel set $A\in\mathcal B(\mathbb R_+^2)$ with finite Lebesgue measure $|A|$, $W(A)$ is a  normal random variable of mean zero and variance $|A|$ and  if $A_1,\cdots,A_n$ are disjoint Borel sets in $\mathcal B(\mathbb R_+^2)$, then $W(A_1),\cdots,W(A_n)$  are mutually independent. We denote by $N$ the Poisson random measure on $\mathbb R_+^3$ with {\it intensity} $\lambda$ which  is a Borel measure on $\mathbb R_+^3$ defined as the product of the Lebesgue measure on $\mathbb R_+\times\mathbb R_+$ with a Borel measure $\mu$ on $\mathbb R^+$ such that $\int_0^{\infty} (z \wedge z^2)\mu(dz)<+\infty$. Note that $\mu$ is a L\'evy measure since $\int_0^\infty (1 \wedge z^2)\mu(dz)<+\infty$.
Recall that for each Borel set $B \in\mathcal B(\mathbb R_+^3)$ with $\lambda(B)<+\infty$, the random variable $N(B)$ has the Poisson distribution with parameter $\lambda(B)$. Moreover, if $B_i, i=1,\ldots, n$ are disjoint Borel sets in $\mathcal B(\mathbb R_+^3)$, then $N(B_1),\cdots,N(B_n)$ are mutually independent. We let $\widetilde N=N-\lambda$ be the compensated Poisson random measure on $\mathbb R_+^3$ associated to $N$.

We introduce the filtration $\mathbb F=(\mathcal F_t)_{t\geqslant 0}$ as the natural filtration generated by the L\'evy random field (see Dawson and Li \cite{DawsonLi}) and satisfying the usual conditions, namely, for any Borel subset $A\in\mathcal B(\mathbb R_+)$ and $B\in\mathcal B(\mathbb R_+^2)$ of finite Lebesgue measure, the processes $(W([0,t]\times A),t\geq 0)$ and $(\widetilde N([0,t]\times B),t\geq 0)$ are $\mathbb F$-martingales.

We consider  the following stochastic differential equation in the integral form. Let $a, b,\sigma, \gamma \in \mathbb R_+$ be constant parameters. Consider the equation:
\begin{equation}
\label{lambda-integral}
	Y(t) = Y(0) + \int_0^t a \left( b  - Y(s)  \right) ds + \sigma \int_0^t \int_0^{Y(s)} W(ds,du)
	+ \gamma \int_0^t \int_0^{Y(s-) } \int_{\mathbb{R}^+} z \widetilde{N} (ds,du, dz),
\end{equation}
where $W(ds,du)$ is a white noise on $\mathbb{R}_+^2$  with unit covariance, $\widetilde N(ds,du,dz)$ is an independent compensated Poisson random measure on $\mathbb{R}_+^3$ with intensity $ds \  du \ \mu(dz)$ with $\mu(dz)$ being a L\'evy measure on $\mathbb{R}_+$ and satisfying
$\int_0^\infty (z \wedge z^2)\mu(dz)<\infty $.

The integrals appearing in Equation \eqref{lambda-integral} (and in the following) are both in the sense of Walsh \cite{Walsh1980}.
It follows from  Dawson and Li \cite[Theorem~3.1]{DawsonLi} or Li and Ma \cite[Theorem~2.1]{LM13} that Equation \eqref{lambda-integral}  has a unique strong solution.

Our model actually belongs to the family of CBI processes. Continuous Branching Processes with Immigration (CBI) are a class of stochastic processes commonly used in modelling population dynamics as in Padoux \cite{Pardoux2016}. The self-exciting features, arising from the integrals in Equation \eqref{lambda-integral} extended on the domain $[0,Y(s))$ with respect to the integration variable $u$, describe the growth of the population due to the reproduction of the previous generations. In the present modelling framework they just describe jumps generated by previous jumps.
We briefly recall the definition by Kawazu and Watanabe \cite{KaW71} [Def. 1.1]. A Markov process $Y$ with state space $\mathbb R_+ $ is called a CBI process characterized by branching mechanism $\Psi(\cdot)$ and immigration rate $\Phi(\cdot)$, if its characteristic representation is given, for $p\geq 0$, by:
\begin{equation}\label{laplace}
\mathbb{E}_{y} \left[e^{- p Y(t)}\right]=\exp\left(-yv(t,p)-\int_{0}^{t}\Phi\big(v(s,p)\big)ds\right),
\end{equation}
where $\mathbb{E}_{y}$ denotes the conditional expectation with respect to the initial value $Y(0) = y$. The function $v:\mathbb R_+\times\mathbb R_+\rightarrow\mathbb R_+$ satisfies the following differential equation:
\begin{equation}\label{ODE0}
\frac{\partial v(t,p)}{\partial t}=-\Psi(v(t,p)),\quad v(0,p)=p
\end{equation}
and $\Psi$ and $\Phi$ are functions of the variable $q\geq 0$ given by
\begin{eqnarray*}
\Psi(q) = a q+\frac{1}{2}\sigma^{2}q^{2}+ \gamma \int_{0}^{\infty}(e^{-qu}-1+qu)\pi(du), \\
\Phi(q) = ab q+\int_{0}^{\infty}(1-e^{-qu})\nu(du),
\end{eqnarray*}
with $\sigma, \gamma \geq 0$, $\beta \in \mathbb{R}$ and  $\pi$, $\nu$ being two L\'evy measures such that
\begin{equation}\label{levy measure moment}
\int_{0}^{\infty} (u\wedge u^{2})\pi(du)<\infty,\quad \int_{0}^{\infty} (1\wedge u) \nu(du)<\infty.
\end{equation}
It is proved in Dawson and Li \cite[Theorem 3.1]{DawsonLi} that the process in Equation  \eqref{lambda-integral} is a CBI process with the branching mechanism $\Psi$ given by:
\begin{equation}\label{equ: Psi general}
\Psi(q)=aq +\frac{1}{2}\sigma^2q^2+ \int_0^{\infty}(e^{-q \gamma z}-1+q \gamma z){\mu}(dz)
\end{equation}
and the immigration rate $\Phi(q)=a b q$.

The link between CBI processes and the affine term structure models has been established by Filipovi\'c \cite{F01}. If the process $Y$ takes values in $\mathbb{R}_+$ he proves equivalence between the two classes. We recall that the joint Laplace transform of a CBI process $Y$ and its integrated process, which is given in Filipovic \cite[Theorem 5.3]{F01}, is defined as follows: for non-negative real numbers $\xi$ and $\theta$, we have:
\begin{equation}\label{non conservative}
{\mathbb E}_y\Big[e^{-\xi Y(t)-\theta\int_0^t Y(s) ds}\Big]
=\exp\Big\{-y v(t,\xi,\theta)-\int_0^t \Phi\big(v(s,\xi,\theta)\big)ds\Big\},
\end{equation}
where $v(t,\xi,\theta)$ is the unique solution of
\begin{equation}\label{ODE1}
\frac{\partial v(t,\xi,\theta)}{\partial t}=-\Psi(v(t,\xi,\theta))+\theta, \quad v(0,\xi,\theta)=\xi.
\end{equation}

\subsection{Hawkes Processes} \label{Hawkes Processes}

A Hawkes process is a special counting process with a random intensity function.
We introduce now the Hawkes processes with exponential kernel. They can be written as follows :
\begin{equation} \label{Hawkes Equation}
Y(t)= Y(0) + \sum_i^{N_t} Z_i = Y(0) + \int_0^t \int_0^{\infty} z J(ds,dz)= ,
\end{equation}
where the last term is an Ito integral, $N_t $ is the number of jumps in the interval between $0$ and $t$  and  $J(dz,ds)$ is a Poisson random measure with intensity $\lambda (t)$, satisfying the SDE:
\begin{eqnarray} \label{Hawkes Intensity}
\lambda (t) &=& \lambda(0) - \beta \int_0^t \lambda (s) ds + \alpha \int_0^t \int_0^\infty z J(ds,dz) \nonumber \\
&=& \exp {(-\beta t)} \lambda(0) + \alpha \sum_i^{N_t}  \exp {[-\beta (t-t_i )} Z_i
\end{eqnarray}
Here $\beta>0$ is the rate of exponential decay of the influence of previous jumps on the intensity level and $\alpha$ the amplitude of the memory kernel, $t_i$ are the jumps times and $Z_i $ the jump sizes, which we shall assume distributed according to an exponential density with parameter $\delta$, so that only positive jumps appear in both Equations \eqref{Hawkes Equation} and \eqref{Hawkes Intensity}, and we can write $\tilde{J}(ds,dz) = J(ds,dz)- \lambda (s) \mu (dz)ds $ and $\mu (dz) = \delta \exp{(- \delta z)} dz$, where $\tilde{J}(ds,dz)$ denotes the compensated version of the Poisson measure $J(ds,dz)$. We assume the following condition holds: $\beta - \alpha/\delta > 0$, granting the non-explosiveness of the Hawkes process (see e.g. Bernis et al. \cite{BerSalSco}).

Hawkes processes with exponential kernel are the only class of Hawkes processes exhibiting both the Markov property and an affine structure (see e.g. Errais et al.\cite{ErraisGieseckeGoldberg}). The have been extensively used in order to describe the dynamics of several asset classes, including equities as in Hainaut and Moraux \cite{HainautMoraux2019}, commodities as in Eyjolfsson and Tj{\o}steim \cite{Eyjolfsson2018}, exchange rates as in Rambaldi et al. \cite{Rambaldi2015} and credit risk as in Errais et al. \cite{ErraisGieseckeGoldberg}.

\section{Forward Prices Modelling}  \label{Forward Model}
In this section we are going to introduce the two alternative models for the forward prices, that we are going to test against electricity market data. In both cases the price at time $t$ of a forward contract with maturity $T \ge t$ is additive and it can be defined as follows
\begin{equation}\label{forward-dynamics}
f(t,T) = \Lambda (t) - \Lambda (0) + \sum_i^n X_i (t,T),
\end{equation}
where $ \Lambda (t)$ is a deterministic seasonality function that will be made precise later on, $n$ is the number of factors used and each of the terms $X_i $ is an underlying factor, whose dynamics will be specified in the following Subsections \ref{CBI_Forward} and \ref{Hawkes_Forward}.

\subsection{The Forward Model based on CBI} \label{CBI_Forward}
Our first model assumes the following dynamics for the factors $X_i, i=1,\dots,n$:
\begin{eqnarray}
X_i (t,T) = X_i (0,T) - \sum_i^n \int_0^t a_i X_i (s) ds + \sigma_i \int_0^t \int_0^{X_i (s,T)} W_i (ds,du) + \nonumber \\
+ \gamma _i \int_0^t \int_0^{X_i (s-,T) } \int_{\mathbb{R}^+} z \widetilde{N}_i (ds,du, dz). \label{factor-dynamics}
\end{eqnarray}
Namely, the $X_i$'s evolve in time with respect to the historical measure $\mathbb{P} $ according to Equation \eqref{lambda-integral} with immigration rate $b_i =0$. By recalling that the intensity of the Poisson random measure $\widetilde{N}_i (ds,du, dz)$ is given by $dsdu\mu _i (dz)$, we assume $\mu _i (dz)= \delta _i \exp {(-\delta _i z )} dz$ with $\delta _i > 0$, for $i=1,\cdots , n$, $z>0$.

It is possible to re-write Equation \eqref{forward-dynamics} as follows:
\begin{eqnarray*}
f(t,T) &=& \Lambda (t) - \Lambda (0) + \sum_i^n X_i (0,T) - \sum_i^n \int_0^t a_i X_i (s) ds + \sum_i^n \sigma_i \int_0^t \int_0^{X_i (s,T)} W_i (ds,du) + \\
&&+ \sum_i^n \gamma _i \int_0^t \int_0^{X_i (s-,T) } \int_{\mathbb{R}^+} z \widetilde{N}_i (ds,du, dz),
\end{eqnarray*}
or, equivalently, as
\begin{eqnarray}
f(t,T) &=& \Lambda (t) - \Lambda (0) +  f(0,T) -  \sum_i^n \int_0^t a_i X_i (s) ds + \sum_i^n \sigma_i \int_0^t \int_0^{X_i (s,T)} W_i (ds,du) + \nonumber \\
&&+ \sum_i^n \gamma _i \int_0^t \int_0^{X_i (s-,T) } \int_{\mathbb{R}^+} z \widetilde{N}_i (ds,du, dz) , \label{eq_f_CBI}
\end{eqnarray}
where $f(0,T)= \sum_{i=1}^n X_i (0,T) $.

\bigskip

The relation between the dynamics of the forward price with respect to the historical measure $\mathbb{P} $ and the risk-neutral dynamics, written with respect to $\mathbb{Q} $, can be easily obtained by applying the following result, proved in the paper by Jiao et al. \cite[{Proposition 4.1}]{JMSS}.

\bigskip

\begin{Proposition}\label{pro:changementprob} Let $X_1, X_2,\cdots, X_n$ be independent CBI processes where for each $i\in\{1,\cdots, n\}$,  $X_i$ is a CBI process under the probability measure $\mathbb{P} $, with dynamics given by Eq. \ref{factor-dynamics}. Assume that the  filtration $\mathbb F=(\mathcal F_t)_{t\geq 0}$
is generated by the random fields $W_1, W_2,\cdots, W_n$ and $\widetilde N_1, \widetilde N_2, \cdots, \widetilde N_n$.  For each $i$, fix
$\eta_i\in\mathbb{R}$ and $\xi_i\in\mathbb{R}_+$ and define
\begin{equation}
U_t:= \sum_i^n \eta_i \int_0^t\int_0^{X_i (s)}W_i(ds,du)+ \sum_i^n \int_0^t \int_0^{X_i (s-)}\int_0^\infty (e^{-\xi_i z }-1)\widetilde{N}_i (ds,du,dz).
\end{equation}
Then the Dol\'eans-Dade exponential $\mathcal{E} (U)$ is a martingale under $\mathbb{P}$ and the probability measure $\mathbb{Q}$ defined by
\begin{equation}
\left. \frac{d\mathbb{Q}}{d\mathbb{P}}\right|_{\mathcal{F}_t}=\mathcal{E} (U)_t,
\end{equation}
is equivalent to $\mathbb{P}$. Moreover, under $\mathbb{Q}$, $X_i$ is a CBI process with parameters $(a_i ^{\mathbb{Q}},b_i ^{\mathbb{Q}},\sigma_i ^{\mathbb{Q}},\gamma_i^{\mathbb Q},\mu_i ^{\mathbb{Q}})$, where:

\begin{eqnarray}\label{par_measure_change}
& a_i^{\mathbb{Q}} =a_i^{\mathbb{P}} -\sigma_i^{\mathbb{P}} \eta_i - \int_0^\infty z(e^{-\theta_iz}-1)\mu_i^{\mathbb{P}} (dz),  & \\
& b_i^{\mathbb{Q}}=a_i^{\mathbb{P}} b_i^{\mathbb{P}} /a_i^{\mathbb{Q}} , \quad  \sigma_i^{\mathbb{Q}}=\sigma_i^{\mathbb{P}}, \quad \gamma_i^{\mathbb{Q}}=\gamma_i^{\mathbb{P}} & \\
& \mu_i^{\mathbb{Q}}(dz)= e^{-\theta_i z} \mu_i^{\mathbb{P}} (dz), \quad \delta _i ^{\mathbb{Q}} = \delta _i ^{\mathbb{P}}
\end{eqnarray}

\end{Proposition}

\begin{Remark}
In this context, the parameters $\eta_i  , \xi_i  $ can be interpreted as the Market Price of Risk associated with the diffusion/jump part of $X_i, i=1,\dots,n$, respectively.
\end{Remark}

\begin{Remark}
In order to avoid arbitrage opportunities we shall assume that the de-seasonalized dynamics of every factor $X_i $ is a local martingale under $\mathbb{Q} $ and this will automatically imply that $a_i= 0$ under $\mathbb{Q} $. Since the first integral is defined with respect to the Gaussian white noise $W_i (ds,du)$ and the second integral is defined with respect to the compensated Poisson random measure $\widetilde{N}_i (ds,du, dz) $, each process $X_i (t,T) $ is in fact a local martingale with respect to $\mathbb{Q} $.
\end{Remark}

\begin{Remark}
From \eqref{par_measure_change}, specifying the relations between the model parameters under the risk-neutral measure $\mathbb{Q} $ and the historical measure $\mathbb{P} $, it is clear that in the present modelling framework, for each factor $X_i $, a mean reversion speed coefficient $a_i$ can be non-null under $\mathbb{P} $ and zero under $\mathbb{Q} $. As far as the immigration term $b_i $ is concerned, if it vanishes under $\mathbb{Q} $, it will be zero under any equivalent probability measure.
\end{Remark}

\begin{Assumption}
In the estimation procedure applied to the real market data we shall assume that only one process of the type introduced in Equation \eqref{factor-dynamics} will drive the forward curve dynamics.
\end{Assumption}

\subsection{The Forward Model Based on Hawkes Processes} \label{Hawkes_Forward}

As alternative to the model proposed in the previous subsection, we consider,under $\mathbb{P} $, Equation \eqref{factor-dynamics} for the instantaneous forward price, where now each $X_i, i=1,\dots, n$ satisfies a SDE of the following form:
\begin{equation}\label{eq_f_H}
X_i (t,T) =  X_i (0,T) - \int_0^t c_i  X_i (s, T) ds + \int_0^t  \sigma_i \sqrt{X_i (s, T)} dW_i (s) + \int_0^t \int_0^\infty z \tilde{J} _i (dz,ds),
\end{equation}
where $\tilde{J} _i (dz,ds)$ are compensated marked point process with intensity $\lambda _i (t)$, satisfying the SDE:
\begin{equation}
\label{Hawkes Intensity2}
\lambda _i (t)= \lambda _i(0) - \beta _i \int_0^t \lambda _i (s) ds + \alpha _i \int_0^t \int_0^\infty z J_i (ds,dz).
\end{equation}

We assume the jump size distributed according to an exponential density with parameter $\delta _i$ for each $(\lambda _i , X_i )$, so we can write:
\begin{equation}
\tilde{J} _i (ds,dz) = J_i (ds,dz) - \lambda _i (s) \mu (dz) ds = J_i (ds,dz) - \lambda _i (s) \delta _i \exp{(- \delta _i z)} (dz) ds.
\end{equation}

\begin{Remark}
The choice of a square-root process for the diffusion part of the forward curves dynamics is motivated by the positivity requirement as well as the choice of the exponential distribution for the jumps size.
\end{Remark}

In order to make the presentation of the two model classes more homogeneous, we can introduce the Dawson-Li representation for the Hawkes-type dynamics as well and write the SDE governing the dynamics of forward prices under the historical measure $\mathbb{P}$ as follows:
\begin{eqnarray}\label{Hawkes Dawson}
X_i (t,T) &=& X_i (0,T) - \int_0^t c_i  X_i (s, T) ds + \int_0^t \int_0^{X_i (s,T)} \sigma _i W_i(du,ds) \nonumber  \\
& & +\int_0^t \int_0^{X_i (s_{-}, T)} \int_{\mathbb R^+} z \tilde{N} _i (dz,du,ds),
\end{eqnarray}
where the definition of the integrals and the notations are the same as in Subsection \ref{CBI} and the $\lambda _i (t)$ evolve according to Eq. \eqref{Hawkes Intensity2}.

It is immediate to remark that the dynamics described by the two model classes look almost identical when written in the Dawson-Li representation, the main difference being the specification of the equation governing the evolution of the intensity processes. This is one of the reasons behind the choice of these two alternative models to describe the forward prices' evolution.

The dynamics just described is given with respect to the historical probability measure $\mathbb{P}$. In order to obtain a description with respect to the risk-neutral measure $\mathbb{Q}$ we need to introduce a measure change. The following proposition provides a measure change preserving the Hawkes-type dynamics. A proof can be found in Bernis et al. \cite{Bernis_Scotti_Sgarra}.

\begin{Proposition}\label{changeP_Hawkes}
Let $(\lambda _i, X_i )$ be described by Equations \eqref{Hawkes Intensity} and \eqref{Hawkes Dawson} under the historical probability $\mathbb{P}$.
Fix $(\eta, \xi) \in  \mathbb{R} \times (-\delta _i, \infty)$  and define:

\begin{equation*}
U_t:= \sum_i^n \eta _i \sigma _i \int_0^t \int_0^{X_i (s)} W_i (ds,du)+ \sum_i^n
\int_0^t \int_0^{\lambda_i (s-)} \int_{\mathbb{R}^+}  \left(e^{-  \xi _i z } -1 \right)\, \widetilde{J}_i (ds,du,dz)
\end{equation*}

Then the Dol\'eans-Dade exponential  $\mathcal{E}(U)$ is a martingale under $\mathbb P$ and the probability measure $\mathbb{Q}$ defined
by $ \left. \frac{d\mathbb{Q}}{d\mathbb{P}} \right|_{\mathcal{F}_t} :=\mathcal{E}(U)_t $ is equivalent to $\mathbb{P}$.
The dynamics with respect to $\mathbb{Q}$ takes the following form:
\begin{eqnarray*}
X_i (t, T) &=& X_i (0,T) + \int_0^t \int_0^{X_i (s,T)} \sigma _i^{\mathbb{Q}} W_i(du,ds) + \int_0^t \int_0^{X_i (s_{-}, T)} z \tilde{J} _i^{\mathbb{Q}} (dz,du,ds), \\
\lambda _i (t) &=& \lambda _i (0) - \int_0^t \beta _i^{\mathbb{Q}} \lambda (s) ds + \alpha _i^{\mathbb{Q}} \int_0^t \int_0^\infty \exp {[-\beta^{\mathbb{Q}} (t-s)]} {J}_i^{\mathbb{Q}}(dz,ds),
\end{eqnarray*}

where

\begin{eqnarray*}
& c_i^{\mathbb{Q}} = c_i^{\mathbb{P}} - \sigma_i^{\mathbb{P}} \eta_i - \int_0^\infty z (e^{-\theta_i z}-1) \mu_i^{\mathbb{P}} (dz), \quad \sigma_i^{\mathbb{Q}}= \sigma_i^{\mathbb{P}} &\\
& \alpha_i^{\mathbb{Q}}= \alpha_i^{\mathbb{P}} , \quad \beta _i^{\mathbb{Q}} = \beta _i^{\mathbb{P}}, \quad
\mu_i^{\mathbb{Q}}(dz)= e^{-\theta_i z} \mu_i^{\mathbb{P}} (dz).&
\end{eqnarray*}

\end{Proposition}

\begin{Remark}
In this context, the parameters $\eta _i , \xi _i $ can be interpreted as the Market Price of Risk associated with the diffusion/jump part of the $i-th $ factor $X_i$, respectively.
\end{Remark}

\begin{Remark}
We shall assume, as for the previous model, that the de-seasonalized dynamics of $X_i $ is a local martingale under $\mathbb{Q} $ and this will automatically imply that the mean reversion speed $c_i $ of any $X_i$ must vanish under $\mathbb{Q} $. Both the diffusion and the jump terms are in fact local martingales with respect to $\mathbb{Q} $.
\end{Remark}

\begin{Remark}
From the formulas in the previous lines, specifying the relations between the model parameters under the risk-neutral measure $\mathbb{Q} $ and the historical measure $\mathbb{P} $, it is clear that in the Hawkes modeling framework, for each factor $X_i $, a mean reversion speed coefficient $c_i$ can be nonzero under $\mathbb{P} $ and zero under $\mathbb{Q} $. A non zero mean-reverting term can then appear in the dynamics written with respect to the historical measure $\mathbb{P} $, although this term vanishes under $\mathbb{Q} $.
\end{Remark}

\begin{Assumption}
In the estimation procedure applied to the real market data we shall assume that only one process of the type introduced in Equation \eqref{factor-dynamics} will drive the forward curve dynamics.
\end{Assumption}

\section{The Futures Dynamics} \label{Futures-Dynamics}
We focus here rigorously on forward contracts delivering a quantity of energy over a finite period of time. We shall refer to them as futures, even if in the literature they are sometimes called swaps or flow forwards.

\begin{Definition}\label{def:future}
The price at time $t \ge 0$ of a futures contract with delivery period $[T_1,T_2]$ with $t \leq T_1 \le T_2$ si given by
\begin{equation}
\label{Futures price}
F(t, T_1 , T_2 ) = \frac{1}{T_2 - T_1} \int_{T_1}^{T_2} f(t, x) \, dx,
\end{equation}
where $f(t,\cdot)$ is the price at time $t$ of the forward contract to be paid upon delivery.
\end{Definition}

\begin{Remark}
	From the above Definition \ref{def:future} it is clear why futures are sometimes called flow forwards: the owner of a futures with delivery period over $[T_1,T_2]$ would substantially receive a constant flow of the commodity over this period. Notice also that a futures contract delivering the commodity over a time period which collapses into a single point coincides with a forward.
\end{Remark}

The value at time $t$ of a Futures contract with delivery period $[T_1 , T_2]$ is given, in our modelling framework, by (recall Equation \eqref{forward-dynamics}):
\begin{equation}
F(t, T_1 , T_2 ) = \frac{1}{T_2 -T_1}  \int_{T_1}^{T_2} f(t,x) dx = \frac{1}{T_2 -T_1} \left[ (\Lambda (t) - \Lambda (0)) + \sum_i^n \int_{T_1}^{T_2} X_i (t,x) dx \right].
\end{equation}
By introducing the dynamics of the factors $X_i$ into the above equation, we get the following equation describing the futures' dynamics under the risk-neutral probability $\mathbb{Q}$ both in the CBI framework (recall Equation \eqref{eq_f_CBI}):

\begin{eqnarray*}
F(t, T_1 , T_2 ) &=& \frac{1}{T_2 -T_1}(\Lambda (t) - \Lambda (0)) + \frac{1}{T_2 -T_1} \int_{T_1}^{T_2} f(0,x) dx + \\
& & + \frac{1}{T_2 -T_1} \sum_i^n \sigma_i \int_{T_1}^{T_2} \int_0^t \int_0^{X_i (s,x)} W_i (ds,dy) dx \\
& & +\frac{1}{T_2 -T_1} \sum_i^n \gamma _i \int_{T_1}^{T_2} \int_0^t \int_0^{X_i (s-,x)} \int_{\mathbb{R}^+} z \widetilde{N}_i (ds, dy, d z ) dx.
\end{eqnarray*}

and in the Hawkes setting (recall Equation \eqref{eq_f_H}):

\begin{eqnarray*}
F(t, T_1 , T_2 ) &=& \frac{1}{T_2 -T_1}(\Lambda (t) - \Lambda (0)) + \frac{1}{T_2 -T_1} \int_{T_1}^{T_2} f(0,x) dx  \\
&&+ \frac{1}{T_2 -T_1} \sum_i^n \sigma_i \int_{T_1}^{T_2} \int_0^t \sqrt{X_i (s,x)} dW_i (s) dx \\
&& +\frac{1}{T_2 -T_1} \sum_i^n \int_{T_1}^{T_2} \int_0^t \int_{\mathbb{R}^+} z \tilde{J} _i (dz,ds) dx .
\end{eqnarray*}

\begin{Assumption}
From now on, in view of our numerical analysis, we will assume that one driving factor is sufficient. Namely, we will consider the case $i=1$.
\end{Assumption}

In order to rule out arbitrage opportunities the prices of futures with different delivery periods must satisfy specific time-consistency relations.
In particular, the value of a futures contract with delivery period $[T_1 , T_n ]$ is linked to the values of the contracts with delivery on intervals $[T_{i} , T_{i+1}], i=1, \dots, n-1 $, where $[T_{i} , T_{i+1}]$ represents a partition of the interval $[T_1 , T_n ]$, by the following relation:

\begin{equation}
F(t, T_1 , T_n ) = \frac{1}{T_n - T_1} \sum_{i=1}^{n-1}(T_{i+1} - T_i ) F(t, T_i , T_{i+1} ) .
\end{equation}

The situation is described by the following picture, describing the so called ``Cascade unpacking mechanism":

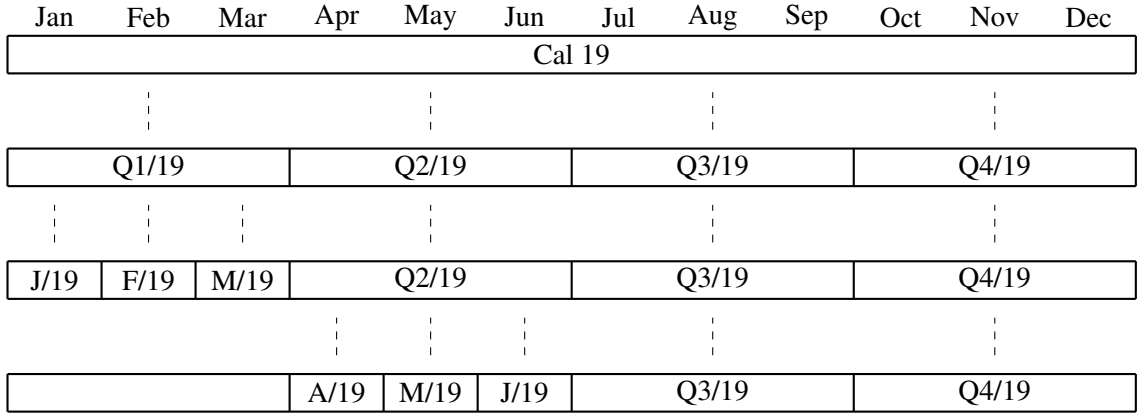
\begin{figure}[htbp]

\centering

\begin{tikzpicture}

\node at (0.625,18.25){Jan};
\node at (1.875,18.25){Feb};
\node at (3.125,18.25){Mar};
\node at (4.375,18.25){Apr};
\node at (5.625,18.25){May};
\node at (6.875,18.25){Jun};
\node at (8.125,18.25){Jul};
\node at (9.375,18.25){Aug};
\node at (10.625,18.25){Sep};
\node at (11.875,18.25){Oct};
\node at (13.125,18.25){Nov};
\node at (14.375,18.25){Dec};

\node at (7.5,17.75){Cal 19};

\draw[thick] (0,18) -- (15,18);
\draw[thick] (0,17.5) -- (15,17.5);
\draw[thick] (0,17.5) -- (0,18);
\draw[thick] (15,17.5) -- (15,18);

\draw[dashed] (1.875,16.75) -- (1.875,17.25);
\draw[dashed] (5.625,16.75) -- (5.625,17.25);
\draw[dashed] (9.375,16.75) -- (9.375,17.25);
\draw[dashed] (13.125,16.75) -- (13.125,17.25);

\draw[thick] (0,16.5) -- (15,16.5);
\draw[thick] (0,16) -- (15,16);
\draw[thick] (0,16) -- (0,16.5);
\draw[thick] (3.75,16) -- (3.75,16.5);
\draw[thick] (7.5,16) -- (7.5,16.5);
\draw[thick] (11.25,16) -- (11.25,16.5);
\draw[thick] (15,16) -- (15,16.5);

\node at (1.875,16.25){Q1/19};
\node at (5.625,16.25){Q2/19};
\node at (9.375,16.25){Q3/19};
\node at (13.125,16.25){Q4/19};

\draw[dashed] (0.625,15.25) -- (0.625,15.75);
\draw[dashed] (1.875,15.25) -- (1.875,15.75);
\draw[dashed] (3.125,15.25) -- (3.125,15.75);
\draw[dashed] (5.625,15.25) -- (5.625,15.75);
\draw[dashed] (9.375,15.25) -- (9.375,15.75);
\draw[dashed] (13.125,15.25) -- (13.125,15.75);

\draw[thick] (0,15) -- (15,15);
\draw[thick] (0,14.5) -- (15,14.5);
\draw[thick] (0,14.5) -- (0,15);
\draw[thick] (1.25,14.5) -- (1.25,15);
\draw[thick] (2.5,14.5) -- (2.5,15);
\draw[thick] (3.75,14.5) -- (3.75,15);
\draw[thick] (7.5,14.5) -- (7.5,15);
\draw[thick] (11.25,14.5) -- (11.25,15);
\draw[thick] (15,14.5) -- (15,15);

\node at (0.625,14.75){J/19};
\node at (1.875,14.75){F/19};
\node at (3.125,14.75){M/19};
\node at (5.625,14.75){Q2/19};
\node at (9.375,14.75){Q3/19};
\node at (13.125,14.75){Q4/19};

\draw[dashed] (4.375,13.75) -- (4.375,14.25);
\draw[dashed] (5.625,13.75) -- (5.625,14.25);
\draw[dashed] (6.875,13.75) -- (6.875,14.25);
\draw[dashed] (9.375,13.75) -- (9.375,14.25);
\draw[dashed] (13.125,13.75) -- (13.125,14.25);

\draw[thick] (0,13.5) -- (15,13.5);
\draw[thick] (0,13) -- (15,13);
\draw[thick] (0,13) -- (0,13.5);
\draw[thick] (3.75,13) -- (3.75,13.5);
\draw[thick] (5,13) -- (5,13.5);
\draw[thick] (6.25,13) -- (6.25,13.5);
\draw[thick] (7.5,13) -- (7.5,13.5);
\draw[thick] (11.25,13) -- (11.25,13.5);
\draw[thick] (15,13) -- (15,13.5);

\node at (4.375,13.25){A/19};
\node at (5.625,13.25){M/19};
\node at (6.875,13.25){J/19};
\node at (9.375,13.25){Q3/19};
\node at (13.125,13.25){Q4/19};

\end{tikzpicture}

\caption{For each given calendar year, as time passes by, forwards are unpacked first in quarters, then in the corresponding months. It may happen that the same delivery period is covered by different contracts, e.g. one simultaneously finds quotes for the monthly contracts Jan/19, Feb/19, Mar/19 and for the quarterly Q1/19.}

\end{figure}


\section{Data Analysis: from Futures Prices to Forward Curves} %
\label{Section: Data Analysis}

From a theoretical point of view the contracts are settled continuously over the delivery period, as you can see from Equation \eqref{Futures price}, but in practice they are  settled at discrete times. Assuming settlement at $N$ points in time $u_{1} < u_{2} < \ldots < u_{N}$, with $u_{1} = T_1$, $u_{N} = T_2$, and $\Delta_{i} = u_{i+1} - u_{i}$, then the discrete version of Equation \eqref{Futures price} becomes

\begin{equation*}
	F(t,T_1, T_2) = \sum_{i=1}^{N}  w(u_{i}, T_1, T_2)f(t,u_{i}) \, \Delta_{i},
\end{equation*}

\noindent where, again, $w(u, T_1, T_2) = \frac{1}{T_2 - T_1}$.

The main goal of what follows is to provide a forward dynamics formulation starting from the futures prices that we observe in the market.
What we are going to do is to build a smooth curve describing today's forward prices from quoted futures prices, according to the Heath-Jarrow-Morton framework outlined before.

This is a well studied problem in literature  and there are basically two approaches to do so: either fitting a parametric function to the entire yield curve by regression, or fitting all observed yields with a spline (see for example Anderson and Deacon \cite{ADA1} for a survey on different methods for constructing yield curves). Here we follow the second approach.\\
Throughout the paper we will also use the notation $T_i^s,T_i^e$ to denote the first (start) and the last (end) day of the delivery period of the $i$-th contract, and $T^s,T^e$ in case of no ambiguity.

Our data set consists of French futures closing prices downloaded from Thomson Reuters, that span over a period of $17$ years, from $2002$ to $2019$.
These contracts are divided with respect to the duration of the delivery period into: weekly (tickers F7B1-B5), monthly (ticker F7BM), quarterly (ticker F7BQ) and yearly (ticker F7BY) contracts. For each of these we have 4 typologies of rolling contracts, namely $c1$, $c2$, $c3$ and $c4$, where $c1$ and $c4$ are the ones with the closest and the farthest delivery period, respectively (for an example to see how rolling contracts work see Section \ref*{Subsection: Working with real data}).\\
On the market we observe the quantity $F(0, T^s , T^e)$ for every contract, for different choices of $T^s , T^e$ ($T^e - T^s = \mbox{7 days}$ for the weekly, $T^e - T^s = \mbox{30 days}$ for the monthly, $T^e - T^s = \mbox{90 days}$ for the quarterly and $T^e - T^s = \mbox{365 days}$ for the yearly), where ``$0$" is the current date, the first available being $\text{July } 1, 2002$, while the last available being $\text{March } 15, 2019$, for a total number of $4234$ current dates.
More precisely, for each day, representing the ``$0$" day, we have a different number of contracts with different delivery periods, depending on the data availability of that day.
We want to extract the curve $f(0,u)$ for all the different choices of the ``$0$" date.
\begin{Notation}
When possible from now on we will write $f(u)$ instead of $f(0,u)$ and $F(T_1 , T_2)$ instead of $F(0, T_1 , T_2)$ to shorten the notation.
\end{Notation}

All the code and the computations have been implemented in MATLAB R2018a,  on a CPU 2.6 GHz and 12 Gb of RAM HP Notebook with Windows $10$.

\subsection{Extracting Smooth Forward Curves from market data}
\label{Section: Extracting a Smooth Forward Curves from market data}

Obtaining a smooth curve of forward prices from futures prices is a well studied problem in the literature, see for example Fleten and Lemming \cite{Fleten03}. The initial condition for using a Heath-Jarrow-Morton approach when modelling forwards is a smooth curve describing today’s forward prices, which must be extracted from the futures prices observed in the market. We will follow the approach by Benth et al. \cite[Ch. 7]{BBK} by imposing the following

\begin{Assumption}
The forward curve can be represented as the sum of two continuous functions $\Lambda(u)$ and $\varepsilon(u)$:
\begin{equation}
\label{Additive Function f}
f(u)= \Lambda(u) + \varepsilon(u), \quad u \in [T^{b},T^{e}],
\end{equation}
where $T^s$ is the starting day of the settlement period for the contract with the closest delivery period and $T^e$ is the last day of the settlement period for the contract with the farthest delivery period.
We interpret $\Lambda(u)$  as a seasonality function and $\varepsilon(u)$ to be an adjustment function
that captures the forward curve's deviation from the seasonality. 

\end{Assumption}
For the specification of the seasonality function we follow Benth et al. \cite{BBK}, namely we define
\begin{equation}
	\label{Seasonality Function}
	\Lambda(u) = a \, \mbox{cos} \left( (u-b) \cdot \frac{2 \pi}{365}\right).
\end{equation}

\noindent The parameter $a \in \mathbb R_+$ is obtained by finding the minimum of the prices over all the contracts, while $b$ is the (normalized\footnote{By normalized distance we mean the distance in days multiplied by 252/365.}) distance between the end of the last day of the year from the day when the minimum occurs.
This procedure leads to:
\begin{equation}
\label{a,b parameters}
	a=13.600 , \quad b=1358.038.
\end{equation}
There are several other methods for extracting the seasonality function from the data (see for example Paraschiv \cite{Paraschiv2013}, and Kiesel et al. \cite{KieselParaschiv2019} for an application to hourly data), but since this topics is not the main focus of our study, we prefer to stick on the well known method proposed by Benth et al. \cite{BKO07} and systematically described in Benth et al. \cite{BBK} (Chap.7, Sect.7.2.1).

We shall see now how the adjustment function $\varepsilon$ is obtained.

\subsubsection{The function $\varepsilon$: a maximum smooth forward curve}
\label{Section: Maximum smooth forward curve}

By following the approach followed by Benth et al. \cite{BBK} and we follow a maximum smoothness criterion applied to the adjustment function $\varepsilon$.
\begin{Remark}
One may ask why the maximum smoothness criterion is applied only to the adjustment function $\varepsilon$ and not to the entire forward function $f$. This ensures the presence of a seasonality pattern that, otherwise, would have possibly been smoothed out.
\end{Remark}
The properties we require for the adjustment function are that it is twice continuously differentiable and horizontal at time $T_{e}$, i.e.
\begin{equation}\label{eq:flatness}
\varepsilon'(T^{e})=0.
\end{equation}
This flatness condition is due to the fact that the long end of the curve may be several years ahead, and obviously the market’s view on risk become less and less sensitive as time goes by.

Let us denote by $C^{2}_0([T^{b},T^{e}])$ the set of real-valued functions on the interval $[T^{b},T^{e}]$ which are twice continuously differentiable with zero derivative in $T_{e}$.
We consider $\mathcal{C}$ as the set of polynomial spline functions of order four which belong to $C^{2}_0([T^{b},T^{e}])$.

\begin{Definition}
We define the smoothest possible forward curve on an interval $[T^{b},T^{e}]$ as the function which minimizes, over $\mathcal C$, the integral
\begin{equation*}
	\int_{T^{b}}^{T^{e}} [\varepsilon'' (u)]^{2} du
\end{equation*}
and such that the closing prices matching condition holds (this is made precise in Equation \eqref{closing price constraint}).
\end{Definition}

We interpret the smoothest forward curve \eqref{Additive Function f} to be the one for which $\varepsilon$ solves the minimization problem above, with $\Lambda$ chosen as in \eqref{Seasonality Function} and $a,b$ as in \eqref{a,b parameters}.

\subsubsection{A smooth forward curve constrained by closing prices}
\label{Subsection:A smooth fwd by cls prices}
In this subsection we present the general procedure  to extract the forward dynamics in a general situation with a fixed number of contracts from the market, but we will often make references to our own case.
Before presenting the algorithm we need to introduce a procedure in order to deal with overlapping periods. Let
\begin{equation*}
	\mathcal{T} = \left\lbrace (T_{1}^{b},T_{1}^{e}), \dots , (T_{m}^{b},T_{m}^{e}) \right\rbrace
\end{equation*}
be a list of start and end dates for the settlement periods of $m$ different futures contracts for a given day (in our case, $m=16$). We need to be able to handle the problem of overlapping settlement periods to rule out arbitrage opportunities. This was a concrete issue working with our data because it happens that, in a given day, two or more contracts have delivery periods that intersect. To overcome this, we construct a new list of dates $\mathcal{\widetilde{T}}$, namely
\begin{equation*}
	\mathcal{\widetilde{T}}=
	\left\lbrace T_{0}, T_{1}, \dots , T_{n} \right\rbrace,
\end{equation*}
where overlapping contracts are split into sub-periods. In our case $n$ is typically $24$, $T_0$ denotes the starting day of the contract with the closest delivery period, while $T_n$ denotes the last day of the contract with the farthest delivery period.
The procedure is illustrated in the following figure:

\begin{figure}[H]
	\centering
	\begin{tikzpicture}
	
	\draw[ very thick, ->] (0,14) -- (14,14);
	
	\draw[color=gray] (1,15) -- (8,15);
	\draw[dashed] (1,14) -- (1,15);
	\draw[dashed] (8,14) -- (8,15);
	\node at (4.5,15.25){settlement period for the first contract};
	\node at (1,13.65){$T_{0}:=T_{1}^{b}$};
	\node at (8,13.65){$T_{2}:=T_{1}^{e}$};
	
	\draw[color=gray] (4,16) -- (12,16);
	\draw[dashed] (4,14) -- (4,15.1) (4,15.45) -- (4,16);
	
	\draw[dashed] (12,14) -- (12,16);
	\node at (8,16.25){settlement period for the second contract};
	\node at (4,13.65){$T_{1}:=T_{2}^{b}$};
	\node at (12,13.65){$T_{3}:=T_{2}^{e}$};
	\end{tikzpicture}
	\caption{Dealing with overlapping delivery time windows.}
	\label{fig:overlapping contracts}
\end{figure}
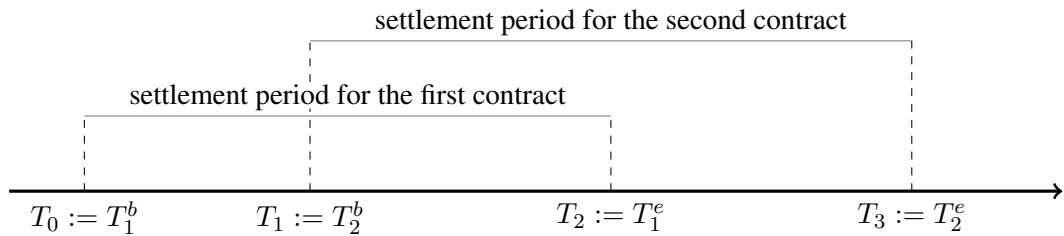

As we can see from Figure \ref{fig:overlapping contracts}, the elements of this new list are basically the elements in $\mathcal{T}$ sorted in ascending order, with duplicate dates removed.
The futures prices could be taken into account either by exact matching or by a constraint on the bid-ask spread prices. Dealing with closing prices, here we impose an exact matching prices on closing prices (see Equation \eqref{closing price constraint}).
From now on we denote with $F_{i}^{C}$ the closing price for the future $i$, $i \in \left\lbrace 1, \dots , m\right\rbrace $.

The adjustment functions $\varepsilon$ is chosen in the class $\mathcal{C}$, namely (with a slight abuse of notation we use $\varepsilon(u;\textbf{\textit{x}})$  instead of $\varepsilon(u) $ to stress the dependence on $\textbf{\textit{x}} $)
\begin{align*}
	\varepsilon(u;\textbf{\textit{x}}) =	
	\begin{cases}
	a_{1} u^{4} + b_{1} u^{3} + c_{1} u^{2} + d_{1} u + e_{1}, \quad u \in [T_{0},T_{1}],\\
	a_{2} u^{4} + b_{2} u^{3} + c_{2} u^{2} + d_{2} u + e_{2}, \quad u \in [T_{1},T_{2}],\\
	\qquad \qquad \quad \; \; \vdots \\
	a_{n} u^{4} + b_{n} u^{3} + c_{n} u^{2} + d_{n} u + e_{n}, \quad u \in [T_{n-1},T_{n}].\\
	\end{cases}	
\end{align*}
where $\textbf{\textit{x}}' = [a_{1}, b_{1}, c_{1}, d_{1}, e_{1}, \dots a_{n}, b_{n}, c_{n}, d_{n}, e_{n}]$ is the row vector of the coefficients of the splines that we want to find.
In this way we have, roughly speaking, a spline for every settlement period.
To find the unknown parameters $\textbf{\textit{x}}' = [a_{1}, b_{1}, c_{1}, d_{1}, e_{1}, \dots a_{n}, b_{n}, c_{n}, d_{n}, e_{n}]$ in order to fully recover the adjustment function, we need to solve the following equality constrained convex quadratic programming problem
\begin{equation}
\label{min-prolem}
	\min_{\textbf{\textit{x}}\in \R^{5n}} \int_{T_{0}}^{T_{n}} [\varepsilon'' (u;\textbf{\textit{x}})]^{2} du,
\end{equation}

\noindent subject to the following constraints:
\begin{itemize}
	\item[i)] continuity of the derivatives up to second order at the knots, for $j = 1, \dots , n-1$,	
	\begin{align}
		\label{knots constraints raw 1}
		a_{j+1}T^{4}_{j} + b_{j+1}T^{3}_{j} + c_{j+1}T^{2}_{j} + d_{j+1}T_{j} + e_{j+1} &= a_{j}T^{4}_{j} + b_{j}T^{3}_{j} + c_{j}T^{2}_{j} + d_{j}T_{j} + e_{j},\\
		\label{knots constraints raw 2}
		4a_{j+1}T^{3}_{j} + 3b_{j+1}T^{2}_{j} + 2c_{j+1}T_{j} + d_{j+1} &= 4a_{j}T^{3}_{j} + 3b_{j}T^{2}_{j} + 2c_{j}T_{j} + d_{j},\\
		\label{knots constraints raw 3}
		12a_{j+1}T^{2}_{j} + 6b_{j+1}T_{j} + 2c_{j+1} &= 12a_{j}T^{2}_{j} + 6b_{j}T_{j} + 2c_{j},
	\end{align}	
	\item[ii)] flatness at the end (see Equation \eqref{eq:flatness})
	\begin{align}
	\label{flatness constraint}
	\varepsilon'(T_{n}; \textbf{\textit{x}}) &= 0,
	\end{align}	
	\item[iii)] matching of the closing prices (see Equation \eqref{Futures price}), for $i = 1, \dots , m$,
	\begin{align}
	\label{closing price constraint}
	F^{C}_{i} &= \frac{1}{T_{i}^{e}-T_{i}^{b}}\int_{T_{i}^{b}}^{T_{i}^{e}} [\Lambda(u) + \varepsilon(u;\textbf{\textit{x}}) ] \, du.
	\end{align}
\end{itemize}

In this way the minimisation problem \eqref{min-prolem} has a total of $3n + m - 2$ constraints (i.e., $3(n-1)$ constraints from \eqref{knots constraints raw 1}-\eqref{knots constraints raw 3}, one constraint from \eqref{flatness constraint} and $m$ constraints from \eqref{closing price constraint}). By computing the second derivative of $\varepsilon$ and inserting it in Equation \eqref{min-prolem} and integrating for every delivery period, we can rewrite the minimisation problem \eqref{min-prolem} as
\begin{equation}
\label{min problem with H}
	\min_{\textbf{\textit{x}}\in \R^{5n}} \textbf{\textit{x}}' \mbox{\textbf{H}} \textbf{\textit{x}},
\end{equation}

\noindent where
\begin{align}
\label{Block Matrix}
	\text{\textbf{H}}= \left[
	\begin{matrix}
	h_{1} & \dots & 0 \\
	& \ddots \\
	0 & \dots & h_{n}
	\end{matrix}
	\right]
	& \quad \text{with} \quad
	h_{j}= \left[
	\begin{matrix}
	\frac{144}{5} \Delta_{j}^{5} & 18 \Delta_{j}^{4} & 8 \Delta_{j}^{3} & 0 & 0 \\
	18 \Delta_{j}^{4} & 12 \Delta_{j}^{3} & 6 \Delta_{j}^{2} & 0 & 0 \\
	8 \Delta_{j}^{3} & 6 \Delta_{j}^{2} & 4 \Delta_{j}^{1} & 0 & 0 \\
	0 & 0 & 0 & 0 & 0 \\
	0 & 0 & 0 & 0 & 0
	\end{matrix}
	\right]	
\end{align}

\noindent and
\begin{equation}
	\label{Delta Coefficients Block Matrix}
	\Delta^{l}_{j} = T_{j}^{l} - T_{j-1}^{l},
\end{equation}
\noindent for $j=1, \dots , n$, and $l=1, \dots , 5$.

We clearly see that the constraints \eqref{knots constraints raw 1}-\eqref{closing price constraint} are linear w.r.t. $\textbf{\textit{x}}$, and so they can be formulated in a matrix form as $\textbf{A\textit{x}} = \textbf{b}$, where $\textbf{A}$ is a $(3n+m-2) \times 5n$-dimensional matrix, and $\textbf{b}$ is a $(3n+m-2)$-dimensional vector. Solving the problem \eqref{min-prolem} with the constraints \eqref{knots constraints raw 1}-\eqref{closing price constraint} is equivalent to solving \eqref{min problem with H} with the constraints written in the form $\textbf{A\textit{x}} = \textbf{b}$.
Let $\lambda′= [\lambda_{1}, \lambda_{2}, \dots , \lambda_{3n+m-2}]$ be the corresponding Lagrange multiplier vector to the constraints \eqref{knots constraints raw 1}-\eqref{closing price constraint}. So, we can now express \eqref{min-prolem} as the following unconstrained minimization problem

\begin{equation}
	\label{lagrange min-problem}
	\min_{\textbf{\textit{x}} \in \R^{5n},\lambda \in \R^{3n+m-2}} \textbf{\textit{x}}'\textbf{H\textit{x}} + \lambda'(\textbf{A\textit{x}} - \textbf{b}).
\end{equation}

\begin{Remark}
	The advantage of dealing with problem \eqref{lagrange min-problem}, instead of \eqref{min-prolem} with the constraints \eqref{knots constraints raw 1}-\eqref{closing price constraint}, is that \eqref{lagrange min-problem} is a unconstrained problem that can be simply solved.
	Indeed the solution $[\bar{\textbf{x}}, \bar{\lambda}]$ is obtained just solving the linear system
	\begin{align}
	\label{equivalent min-problem}
	\left[
	\begin{matrix}
	2\textbf{\textup{H}} & \textbf{\textup{A}}' \\
	\textbf{\textup{A}} & \textbf{\textup{0}}
	\end{matrix}
	\right]
	&
	\left[
	\begin{matrix}
	\textbf{\textit{x}} \\
	\lambda
	\end{matrix}
	\right]
	=
	\left[
	\begin{matrix}
	\textbf{\textup{0}} \\
	\textbf{\textup{b}}
	\end{matrix}
	\right].
	\end{align}
	
	The dimension of the left matrix is $(8n+m-2) \times (8n+m-2)$. 
	Solving \eqref{equivalent min-problem} numerically is standard, and can be done using various techniques (e.g. QR or LU factorisation)\footnote{In this work we have used LU factorisation, since in our case it performed better than the QR ones, i.e. it gives a conditioning number smaller than the QR ones.}.	
\end{Remark}

\subsection{Numerical Results}
\label{Subsection: Working with real data}

Recall that we are working with closing prices of French Futures from $2002$ to $2019$. 
The yearly contracts span from $2002$ to $2019$, the quarterly from $2011$ to $2019$, the monthly from $2011$ to $2019$ and the weekly from $2010$ to $2019$.
We are working with \textit{rolling contracts}, called $c1$, $c2$, $c3$, and $c4$.
Below there is an example which shows how these contracts roll for a monthly contract.

\begin{figure}[H]
	\centering
	\begin{tikzpicture}
	
	\draw[ very thick, ->] (0,14) -- (15,14);
	
	\draw[color=gray] (1.5,15) -- (4,15);
	\draw[color=gray] (4.5,15) -- (7,15);
	\draw[color=gray] (7.5,15) -- (10,15);
	\draw[color=gray] (10.5,15) -- (13,15);
	
	\draw[dashed] (0.25,14) -- (0.25,15);
	\draw[dashed] (1.5,14) -- (1.5,15);
	\draw[dashed] (4,14) -- (4,15);
	\draw[dashed] (4.5,14) -- (4.5,15);
	\draw[dashed] (7,14) -- (7,15);
	\draw[dashed] (7.5,14) -- (7.5,15);
	\draw[dashed] (10,14) -- (10,15);
	\draw[dashed] (10.5,14) -- (10.5,15);
	\draw[dashed] (13,14) -- (13,15);
	
	\node at (0.25,15.25){Start};
	\node at (2.75,15.25){c1 contract};
	\node at (5.75,15.25){c2 contract};
	\node at (8.75,15.25){c3 contract};
	\node at (11.75,15.25){c4 contract};
	
	\node at (0.25,12.8){\rotatebox{-90}{17/01/2014}};
	\node at (1.5,12.8){\rotatebox{-90}{01/02/2014}};
	\node at (4,12.8){\rotatebox{-90}{28/02/2014}};
	\node at (4.5,12.8){\rotatebox{-90}{01/03/2014}};
	\node at (7,12.8){\rotatebox{-90}{31/03/2014}};
	\node at (7.5,12.8){\rotatebox{-90}{01/04/2014}};
	\node at (10,12.8){\rotatebox{-90}{30/04/2014}};
	\node at (10.5,12.8){\rotatebox{-90}{01/05/2014}};
	\node at (13,12.8){\rotatebox{-90}{31/05/2014}};
	
	\end{tikzpicture}
	\caption{An example of how rolling contracts work, in the case when Today, the ``$0$" date, is $\text{January } 17, 2014$.}
	\label{fig:rolling contracts}
\end{figure}
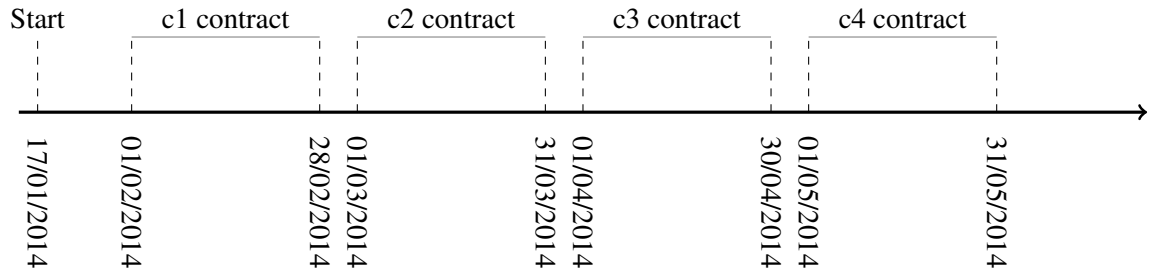

As you can see from Figure \ref{fig:rolling contracts}, the c1 contract is the closer one to the current date and its delivery period spans from 01/02/2014 to 28/02/2014. After 28/02/2014, there is a rollover from the c1 contract to the c2 contract and so on.

The following Figure \ref{fig:subfig1} shows the plot of the futures closing prices for $c1, c2, c3$ and $c4$ contract for the weekly contract.
In the $x$-axis there are the different dates, while in the $y$-axis there is the price. As you can see the presence of seasonality is pretty strong and this could also be seen from the monthly and quarterly contracts.

\begin{figure}[h!]
	\centering
	\subfloat[][\emph{c1} contract]
	{\includegraphics[width=.45\textwidth]{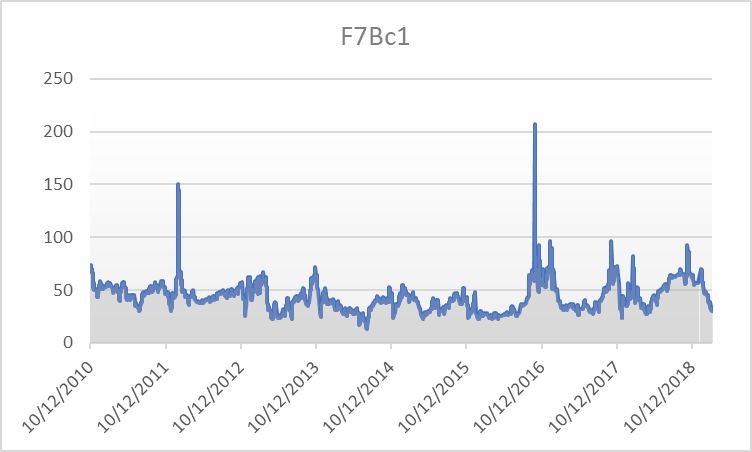}} \quad
	\subfloat[][\emph{c2} contract]
	{\includegraphics[width=.45\textwidth]{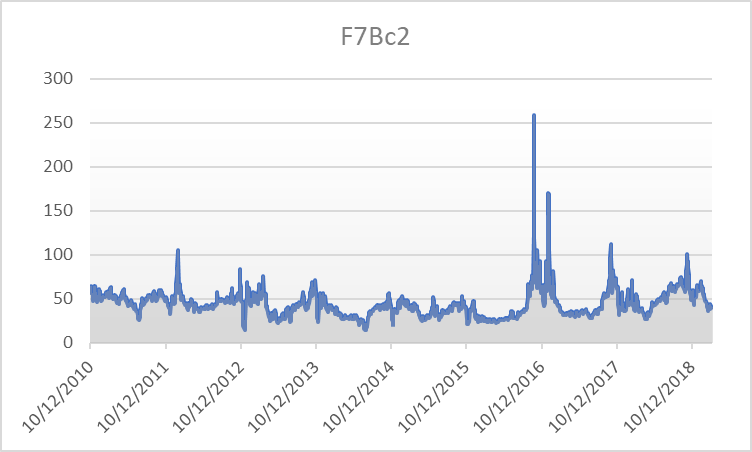}} \\
	\subfloat[][\emph{c3} contract]
	{\includegraphics[width=.45\textwidth]{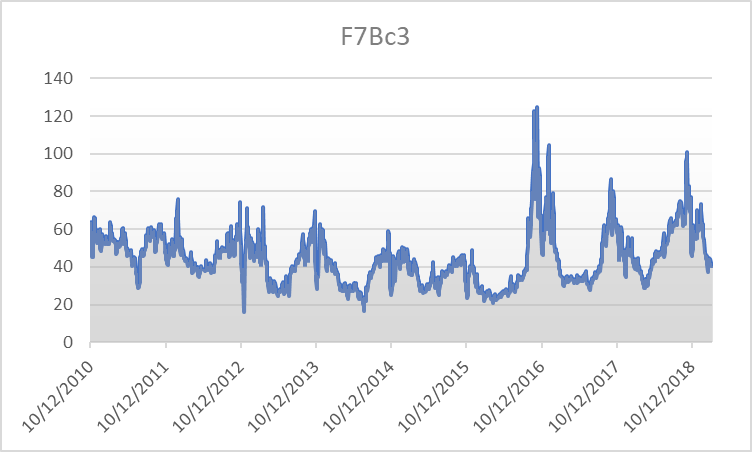}} \quad
	\subfloat[][\emph{c4} contract]
	{\includegraphics[width=.45\textwidth]{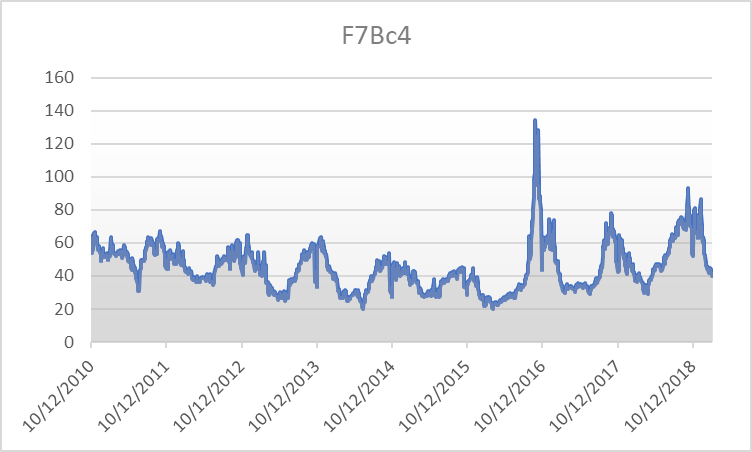}}
	\caption{Weekly prices in Euros of the contracts: the \emph{c1} and \emph{c4} contracts start from $\text{December } 10, 2010$ and end on $\text{March } 14, 2019$, while the \emph{c2} and \emph{c3} contracts start from $\text{December } 10, 2010$ and end on $\text{March } 15, 2019$.}
	\label{fig:subfig1}
\end{figure}

Since we were worried that our analysis could have been affected by the presence of the quarterly contracts being sensible to the seasonality pattern, we did the analysis in both cases, with and without the quarterly contracts.
After the analysis was performed, we noticed that the results were coherent in both cases, so from now on we will focus only on contracts different from quarterly.

The algorithm presented takes approximately $40$ seconds to extract the $4234$ different curves, i.e. the curves $f(0,u)$, for all the $4234$ different values of ``$0$".\\

\begin{figure}[H]
	\hspace*{-2cm}\includegraphics[scale=0.4]{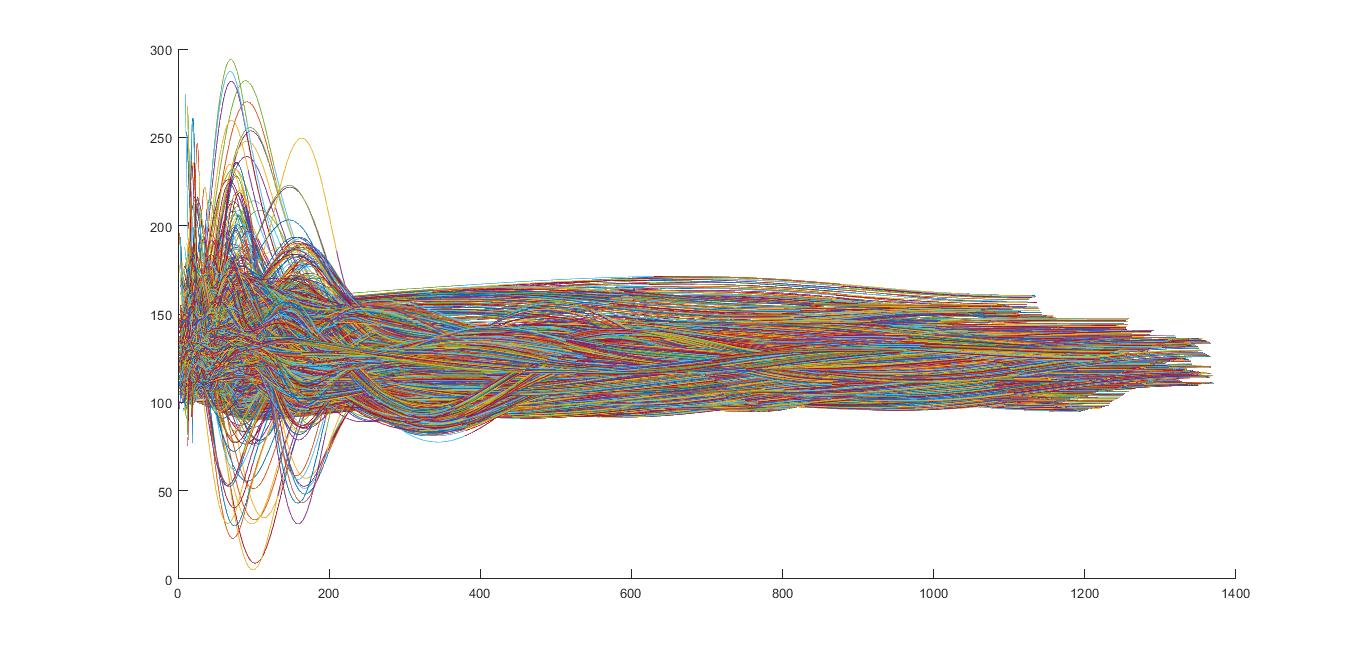}
	\vspace*{-1cm}
	\caption{De-seasonalized forward curves extracted without considering quarterly contracts. In the $x$-axis there is the time to maturity while in the $y$-axis there are the prices in euros.}
	\label{fig:Forward Curves Total}
\end{figure}

Figure \ref{fig:Forward Curves Total} shows, for every day from $2002$ to $2019$, a different curve $f(0,u)$, for a total number of $4234$ curves.
In the $x$-axis we have the time to maturity while in the $y$-axis we have the prices.
Note that different colours in the curves mean different type of contracts.
Note also that the further we move on the $x$-axis, the flatter the curves become. This is in line with the flatness constraint in Equation \eqref{eq:flatness}.
The reason behind the difference in the shapes of the curves is to be investigated in the price constraint but also in the nature of the contracts, since for each day the number and the type of available contracts were different, having to deal also with overlapping settlement periods.

\section{Jump Detection}
\label{Section: Jumps Detection}

We now want to detect the jumps. We will be only dealing with positive jumps since we have supposed that the jump size is distributed according to an exponential density, as already described in Section \ref{Forward Model}.
In the next subsection we will describe an algorithm that allows to detect jumps, and, as a by-product, which also gives their size.

\medskip

\subsection{Description of the Algorithm}
\label{Subsection:JumpsAlgorithm}

In order to detect jumps we proceed in the following way: for a fixed maturity $T$, we define
$$
V_{t} = f(t,T),
$$
the vertical section at maturity $T$, where the parameter $t$ ranges through all the curves, i.e. $t=1, \dots, 4234$.
Roughly speaking, looking at Figure \ref{fig:Forward Curves Total}, this is nothing but the intersection between the vertical line $x=T$ and the curves.
There are several ways to detect jumps from the data, maybe the more natural one consisting in fixing a threshold $\Theta \in \R_{+}$ and saying that a jump occurs at time $\tilde t$ if $|V_{\tilde t+1}-V_{\tilde t}| \geq \Theta$.
We follow here an iterative weighted least square approach. Define $n=4234$ the total number of curves and $\mathcal{N}=\left\lbrace 1,2, \dots , n-1 \right\rbrace $. The algorithm to detect jumps reads as follows:

\begin{enumerate}
	\item Define $\sigma_{1}^2 = \frac{1}{n-2} \sum_{t \in \mathcal{N}} \frac{(V_{t+1} - V_{t})^{2}}{V_{t}}$;
	\item Identify all the $t \in \mathcal{N}$ such that $\frac{V_{t+1} - V_{t}}{\sqrt{V_{t}}} \geq 3\sigma_{1}$ and denote by $\mathcal{M}_1 \subseteq \mathcal{N}$ this family of indices, so that $m_{1}=|\mathcal{M}_1|$; 
	\item Define $\sigma_{2}^2 = \frac{1}{n-m_{1}-1} \sum_{t \in \mathcal{N} \smallsetminus \mathcal{M}_1} \frac{(V_{t+1} - V_{t})^{2}}{V_{t}}$;
	\item Identify all the $t \in (\mathcal{N} \smallsetminus \mathcal{M}_1)$ such that $\frac{V_{t+1} - V_{t}}{\sqrt{V_{t}}} \geq 3\sigma_{2}$ and denote by $\mathcal{M}_2 \subseteq (\mathcal{N} \setminus \mathcal M_1)$ this family of indices, so that $m_{2}=|\mathcal{M}_2|$;
	\item Iterate the procedure updating $\sigma_i^2 = \frac{1}{n-(\sum_{j=1}^{i-1} m_{j}) -1} \sum_{t \in \mathcal{N} \smallsetminus (\bigcup_{j=1}^{i-1} \mathcal{M}_j)} \frac{(V_{t+1} - V_{t})^{2}}{V_{t}}$ and  $\mathcal{M}_i \subseteq \mathcal{N} \smallsetminus (\bigcup_{j=1}^{i-1} \mathcal{M}_j)$; 
	\item Stop when finding $k \in \N$ such that $m_{k}=|\mathcal{M}_k|=0$ (no new jumps are detected).
\end{enumerate}

This procedure finds, at every iterations, new jumps. Clearly as the number of iterations increases, $\sigma_i$ decreases and so the jumps detected become smaller and smaller.
After several tests on the data, we noticed that stopping at $k \in \N$ such that $m_k=0$ would lead to  too many jumps, of which the last detected are much smaller compared to the ones discovered at the first iterations. So we chose, as a good compromise, to stop the algorithm after the first two iterations.

\subsection{Jumps Analysis}

We selected different values of $T$, namely $T=200$, $T=400$ and $T=700$ days. This covers all the different shapes of the forward curves and so it represents a good sampling of our data.
After applying the algorithm described in Subsection \ref{Subsection:JumpsAlgorithm} we end up with the following pictures showing the size and distribution of the jumps detected at $T=200$, represented by the orange vertical lines, together with the corresponding price plot, represented by the continuous blue line:

\begin{figure}[H]
	\begin{minipage}{.5\linewidth}
		\centering
		\subfloat[][Jumps detected at the 1° iteration]{\label{iterazione_1_200}\includegraphics[scale=.38]{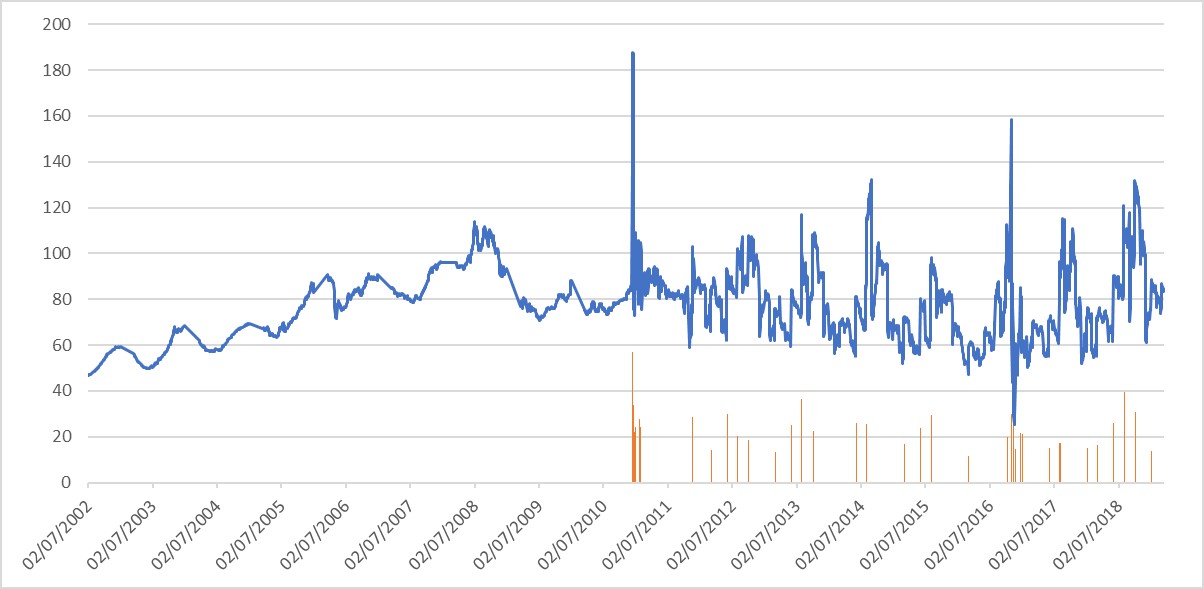}}
	\end{minipage}%
	\begin{minipage}{.5\linewidth}
		\centering
		\subfloat[][Jumps detected at the 2° iteration]{\label{iterazione_2_200}\includegraphics[scale=.38]{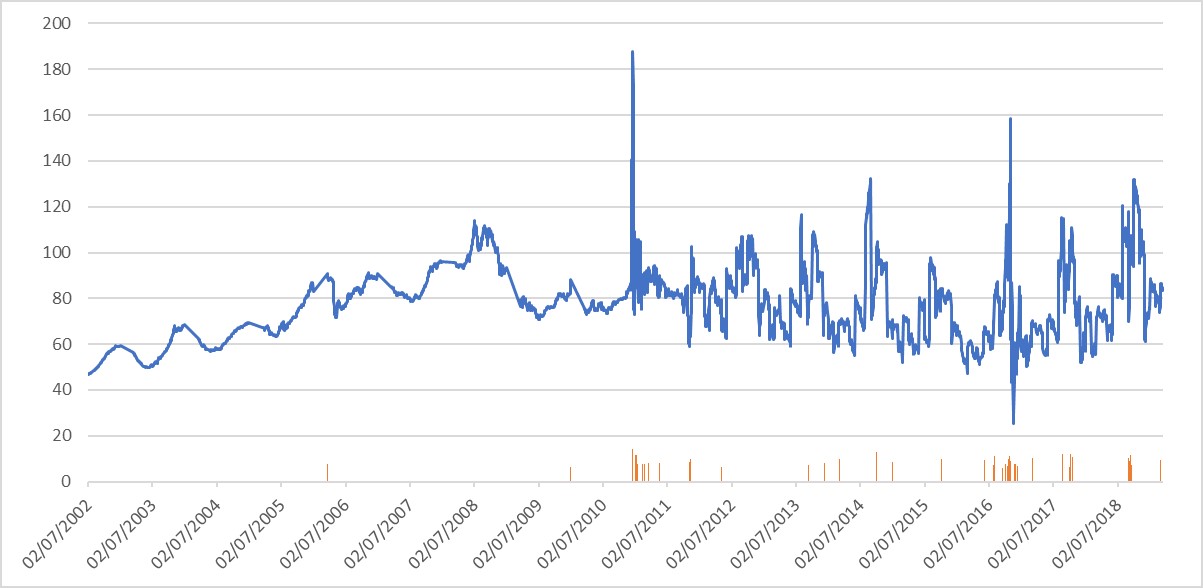}}
	\end{minipage}\par\medskip
	\centering
	\subfloat[][Total jumps detected]{\label{iterazione_1+2_200}\includegraphics[scale=.38]{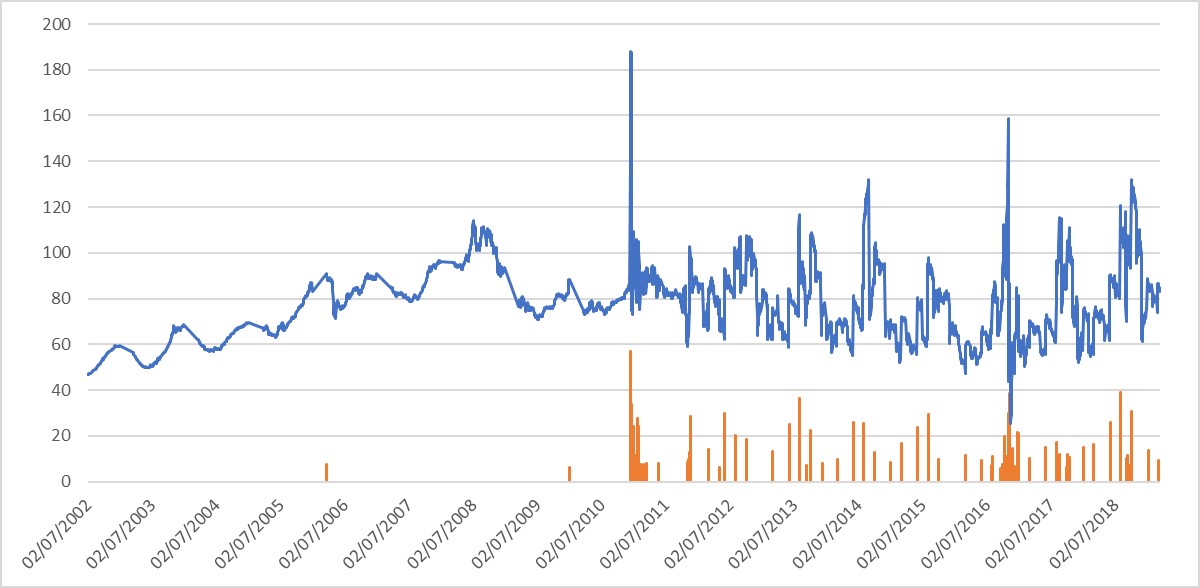}}
	\caption{Jumps detected (vertical orange lines) and price plot (blue line) at $T=200$ at different iterations.}
	\label{jump_price_200}
\end{figure}

The number of detected jumps at the various iterations and for different maturities is listed below:
\begin{table}[H]
	\centering
	\begin{tabular}{ |c|c|c|c| } 
		\hline
		$T$ & \quad \quad $200$ \quad \quad \quad & \quad \quad $400$ \quad \quad \quad & \quad \quad $700$ \quad \quad \quad \\
		\hline
		\hline
		\text{1° iteration} &  38 & 19 & 43 \\
		\hline
		\text{2° iteration} & 48 & 55 & 36 \\
		\hline
		\text{Total} & 86 & 74 & 79 \\
		\hline
	\end{tabular}
	\caption{Number of jumps detected at different maturities and at different iterations.}
	\label{tab: number of jumps}
\end{table}

As one can see from the pictures above, at $T=200$ the jumps detected are the bigger ones with respect to their amplitude, and this is not surprising looking at Figure \ref{fig:Forward Curves Total}, where one can clearly see that the price movements are pretty significant at $T=200$.
On the other hand, when $T=700$, the jumps detected are quite small and this is due to the fact that at $T=700$ the curves are pretty flatten, leading to prices which are close to each other.

\section{Parameters Estimation}

Before starting with the statistical tests on the two models, we still have to estimate: the size of the jumps and the parameters characterizing the drift and the volatility coefficients.
We start with $\delta$, the jumps' size.
Recall that in Sections \ref{Hawkes_Forward} and \ref{CBI_Forward} we assumed for both models the jumps' size to be distributed like an exponential random variable with parameter $\delta >0$. Let $z_i$ be the size of the $i$-th jump, where $i=1,\dots , L$ and $L$ is the number of jumps at the chosen maturity $T$ ( see Table \ref{tab: number of jumps}). Then $\delta$ can be estimated e.g. via its Maximum Likelihood Estimator:
\begin{equation}
\hat{\delta}= \frac{L}{\sum_{i=1}^{n} z_i}.
\end{equation}
We obtain what follows (the fact that we have the smallest values of $\hat{\delta}$ at $T=200$ is not surprising at all, because at the beginning jumps are bigger, as said before):
\begin{table}[H]
	\centering
	\begin{tabular}{ |c|c|c|c| }
		\hline
		$T$ & $200$ & $400$ & $700$  \\
		\hline
		\hline
		$\widehat{\delta}$ & 0.064282682
		& 0.194300598
		& 0.430239733
		\\
		\hline
	\end{tabular}
	\caption{Parameter estimation for $\delta$.}
	\label{tab: delta_estimation}
\end{table}

We now need to estimate the parameter appearing in the drift coefficient of our forward dynamics.
\begin{Remark}
Notice that neither in Equation \eqref{eq_f_CBI} (stated in the CBI framework), nor in Equation \eqref{Hawkes Dawson} (given for the Hawkes case) the mean-reversion term appears, but it does if we pass under the measure $\mathbb P$ by exploiting, respectively, Propositions \ref{pro:changementprob} and \ref{changeP_Hawkes}. In both cases we end up with the following dynamics under $\mathbb P$, for a fixed $T$:
$$
\widetilde X(t+1,T) = \widetilde X(t,T) -  \int_t^{t+1} \widetilde a \widetilde X(s,T) ds
$$
where $\widetilde X$ denotes the factor appearing in the forward dynamics without the seasonality and with no jumps (i.e., we remove all the times at which a jump has occurred).
\end{Remark}
In order to estimate $\widetilde a$ we simply discretize the above equation, by writing
\begin{equation}
\widehat {\widetilde a } = 1 - \frac{\widetilde X(t+1,T)}{\widetilde X(t,T)},
\end{equation}
so that the estimates follow:
\begin{table}[H]
	\centering
	\begin{tabular}{ |c|c|c|c| }
		\hline
		$T$ & $200$ & $400$ & $700$  \\
		\hline
		\hline
		$\widehat{\tilde a}$ & -0.001387342
		& -0.001771845
		& -0.000237952
		\\
		\hline
	\end{tabular}
	\caption{Parameters estimation for $\tilde a$}
	\label{tab: mean_rev_speed_a_estimation}
\end{table}
As you can see from Table \ref{tab: mean_rev_speed_a_estimation}, the estimated value of $\widetilde a$ in all the three cases is really small, very close to $0$.

Now it remains to estimate the volatility parameter $\sigma$, appearing in both Equations \eqref{eq_f_CBI} and in Equation \eqref{Hawkes Dawson}. By recalling the iterative algorithm presented in Subsection \ref{Subsection:JumpsAlgorithm} and taking as $\widehat \sigma$ the value of $\sigma_2$ (namely, the estimation after the second iteration), we get
\begin{table}[H]
	\centering
	\begin{tabular}{ |c|c|c|c| }
		\hline
		$T$ & $200$ & $400$ & $700$  \\
		\hline
		\hline
		$\widehat{\sigma}$ & 0.218667945
		& 0.129560813 & 0.066361067		
		\\
		\hline
	\end{tabular}
	\caption{Parameter estimation for $\sigma$.}
	\label{tab: sigma_estimation}
\end{table}

\section{Testing the Models}
\label{Section: Testing the Models}

In this section we want to perform statistical tests concerning the intensity of the jumps. 
We want to check what is the best process modelling the jumps we have detected before.
We test the two models based on Hawkes and branching processes, plus the Poisson, which is a toy-model:
\begin{itemize}
	\item[(0)] Poisson process;
	\item[(1)] Hawkes process;
	\item[(2)] Self exciting branching process.
\end{itemize}
We will mainly rely on the Kolmogorov-Smirnov (KS) test, namely we will test the null hypothesis $H_0$, stating that the data have the same cumulative distribution function as the one coming from one of the above models, against the alternative $H_1$. We fix a significance level equal to $0.05$.\\

\subsection{Jump Intensity Estimation}

Before using the KS test to check whether the jumps distribution comes from one of the three models, we need to estimate the intensity from our data.
The input in all the cases will be the time occurrences of the jumps over $[0,T]$ (for the three different values of $T$), $0< \tau_{1} < \tau_{2} <  \dots < \tau_{N}=T$, where $N$ can take the values $86$, $74$ and $79$ depending on the chosen maturity $T$, as you can check from Table \ref{tab: number of jumps}.
\begin{itemize}
	\item[(0)] [Poisson]  The (constant) intensity, $\lambda^P>0$, is estimated as the ratio between the total number of (positive) jumps and the sum of the inter-times between two consecutive jumps.
	\item[(1)] [Hawkes] Here the intensity is given by Equation \eqref{Hawkes Intensity2}, so this case will be treated in a separate subsection.
	\item[(2)] [Branching] In this case the stochastic intensity $\lambda^B(t) \propto X(t,T)$ and the constant of proportionality $\gamma$ is estimated, for a fixed $T$, as the ratio between the total number of (positive) jumps and the cumulative (de-seasonalized) forward prices.
\end{itemize}

\subsubsection{The Hawkes Setting: Estimating $\lambda$}
Recalling Equation \eqref{Hawkes Intensity2}, it is clear that we have to estimate three parameters:  $\lambda(0)$, $\alpha$ and $\beta$.
We mainly rely on the paper by Ozaki \cite{Ozaki1979} and we will find a Maximum Likelihood Estimation (MLE).\\
The log-likelihood of a Hawkes process whose response function is of the form $\alpha e^{-\beta t}$, is given by

\begin{equation}
	\label{Log likelihood function}
	\log L\left(\tau_{1}, \cdots, \tau_{N}\right)=-\lambda(0) \tau_{N}+\sum_{i=1}^{N} \frac{\alpha}{\beta}\left(e^{-\beta\left(\tau_{N} -\tau_{i}\right)}-1\right)+\sum_{i=1}^{N} \log \left( \lambda(0)+\alpha A(i) \right) ,
\end{equation}
where
$A(i)=\sum_{\tau_{j} < \tau_{i}} e^{-\beta\left(\tau_{i} - \tau_{j} \right)}$ for $i \geq 2$ and $A(1)=0$.

In order to estimate the parameters $\lambda(0), \alpha,\beta$, we need to find the maximum of the function in \eqref{Log likelihood function}, which is a real value function of three variables. The maximum was found using the command \textit{fminsearch} of Matlab.

The following three tables show the parameters estimated for the jump intensity of the three different models at the maturities $T=200$, $T=400$ and $T=700$. As far as the branching model is concerned, $\lambda^B(t)$ is proportional to the process $X(t,T)$ (for a fixed $T$), and the parameter to be estimated is the $\gamma$, which is the constant ratio between the two processes.

\bigskip

\begin{table}[H]
	\centering
	\begin{tabular}{cp{1.5cm}p{1.5cm}p{1.5cm}p{1.5cm}p{1.5cm}}
		\midrule
		\multicolumn{1}{c}{Model} & \multicolumn{5}{c}{Parameters}\\
		\cmidrule(r){1-1} \cmidrule(l){2-6}
		& $\lambda(0)$ & $\alpha$ & $\beta$ & $\lambda^P$ & $\gamma$ \\
		\midrule
		Poisson & {--} & {--} & {--} & $0.023$ & {--} \\
		\addlinespace
		Hawkes & {0.017} & 0.074 & 0.094 & {--} & {--} \\
        \addlinespace
        Branching & {--} & {--} & {--} & {--} & $0.00028$ \\
		\midrule
	\end{tabular}
	\caption{Parameters estimation for the three models at $T=200$.} \label{tab:parameters_estimation_T=200}
\end{table} 	

\bigskip

\begin{table}[H]
	\centering
	\begin{tabular}{cp{1.5cm}p{1.5cm}p{1.5cm}p{1.5cm}p{1.5cm}}
		\midrule
		\multicolumn{1}{c}{Model} & \multicolumn{5}{c}{Parameters}\\
		\cmidrule(r){1-1} \cmidrule(l){2-6}
		& $\lambda(0)$ & $\alpha$ & $\beta$ & $\lambda^P$ & $\gamma$ \\
		\midrule
		Poisson & {--} & {--} & {--} & $0.018$ & {--} \\
		\addlinespace
		Hawkes & $0.0026$ & $0.012$ & $0.016$ & {--} & {--} \\
		\addlinespace
		Branching &  {--} & {--} & {--} & {--} & $0.00021$ \\
		\midrule
	\end{tabular}
	\caption{Parameters estimation for the three models at $T=400$.} \label{tab:parameters_estimation_T=400}
\end{table}

\bigskip

\begin{table}[H]
	\centering
	\begin{tabular}{cp{1.5cm}p{1.5cm}p{1.5cm}p{1.5cm}p{1.5cm}}
		\midrule
		\multicolumn{1}{c}{Model} & \multicolumn{5}{c}{Parameters}\\
		\cmidrule(r){1-1} \cmidrule(l){2-6}
		& $\lambda(0)$ & $\alpha$ & $\beta$ & $\lambda^P$ & $\gamma$ \\
		\midrule
		Poisson & {--} & {--} & {--} & $0.0019$ & {--} \\
		\addlinespace
		Hawkes & $0.040$ & $0.059$ & $0.085$ & {--} & {--} \\
		\addlinespace
		Branching &  {--} & {--} & {--} & {--} & $0.00032$  \\
		\midrule
	\end{tabular}
	\caption{Parameters estimation for the three models at $T=700$.} \label{tab:parameters_estimation_T=700}
\end{table}

\subsection{KS test for the models}

We perform a Kolmogorov-Smirnov test in order to check which of the proposed distributions best models the jumps in our data. At the end of the subsection we will provide the p-values to conclude.\\

\begin{itemize}
	\item[(0)] [Poisson] We check whether the jumps inter-times are drawn from an exponential distribution with parameter $\lambda^P$, where $\lambda^P$ is the one given in Tables \ref{tab:parameters_estimation_T=200}, \ref{tab:parameters_estimation_T=400} and \ref{tab:parameters_estimation_T=700}. The $p$-value is automatically given by Matlab via the function \textit{kstest}.
	\item[(1)] [Hawkes] We mainly adapt the methods in Lallouache and Challet \cite{LC16} to our purpose. In particular, we check if the time-deformed series of durations $\left\lbrace \theta_i \right\rbrace_{i=1,\dots,N} $, defined by
	\begin{equation}
	\theta_{i}=\int_{\tau_{i-1}}^{\tau_{i}} \widehat{\lambda}_{t} d t,
	\end{equation}
	has an exponential distribution of parameter $1$, where $\widehat{\lambda}_{t}$ is the intensity estimated before (the estimated parameters can be found in Tables \ref{tab:parameters_estimation_T=200}, \ref{tab:parameters_estimation_T=400} and \ref{tab:parameters_estimation_T=700}), and where recall that the $\tau_i$'s are the jumps arrival times.
	The $p$-value is automatically given by Matlab via the function \textit{kstest}.
	\item[(2)] [Branching] The procedure here is quite different from the ones adopted before and it is the object of the following subsection.
\end{itemize}

\subsubsection{Setting the KS Test for the Branching Model}

The KS test we will perform in this case was constructed based on the following classical result.
\begin{Proposition}
	\label{NHPP Proposition}
	Let $({N}_t)_{t \geq 0}$ be a non homogeneous Poisson process with continuous expectation function. 
	If $n$ events have occured in $(0,T]$, then the arrival times $\tau_1, \dots , \tau_n$ are distributed as the order statistics from a sample with cumulative distribution function
	\begin{equation}\label{CDF_CBI}
	F(t) = \frac{\int_{0}^{t} \lambda^B(s) ds}{\int_{0}^{T} \lambda^B(s) ds}, \quad 0 \leq t \leq T,
	\end{equation}
	where in our case $\lambda^B(s) \propto f(s,T)$, for a fixed $T$.
\end{Proposition}

\begin{Remark}
	Recall that $\lambda^B(s) \propto f(s,T)$, for any fixed $T$. It is crucial to notice, from Proposition \ref{NHPP Proposition}, that the distribution of the arrival times is independent of the factor of proportionality connecting $\lambda^B$ and $f(\cdot, T)$.
\end{Remark}

So in this case we perform a KS test, comparing the cumulative distribution function $F(t)$ in Equation \eqref{CDF_CBI} with the empirical one relative to the jump times.\\
Since $F(t)$ given in Proposition \eqref{NHPP Proposition} is not a priori associated to a known distribution, we cannot use the Matlab command kstest and we have to rely on the classical theory on the KS test.
The KS statistics for the test is
\begin{equation}
	D_{n}=\sup _{x \in \R}\left|S_{n}(x)-F(x)\right|,
\end{equation}
where $n$ is the number of our data (recall Table \ref{tab: number of jumps}), $F$ is the cumulative distribution function in Equation \eqref{CDF_CBI} and $S_n(x)$ is the empirical cumulative distribution function of the jump arrival times. We find:
\begin{table}[h!]
	\centering
	\begin{tabular}{ |c|c|c|c| }
		\hline
		$T$ & $200$ & $400$ & $700$ \\
		\hline
		\hline
		$D_n$ & 0.2151 & 0.2276 & 0.2513 \\	
		\hline
	\end{tabular}
	\caption{Maximum distance between the empirical distribution function and the theoretical one for the Branching case.}
\end{table}

In the following Figure \ref{cdf ecdf comparison} we graphically compare the two cumulative distribution functions at the three different maturities.

\begin{figure}[H]
	\begin{minipage}{.5\linewidth}
		\centering
		\subfloat[][T=200]{\label{cdf sez 200}\includegraphics[scale=.38]{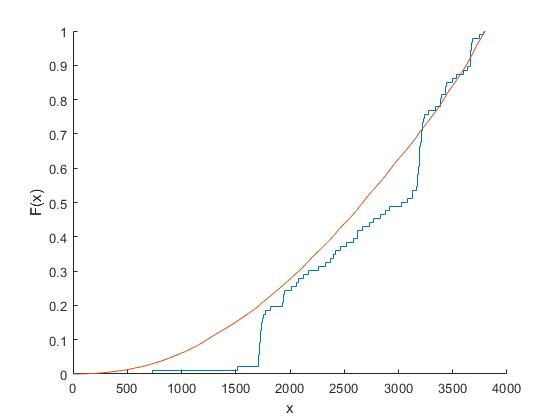}}
	\end{minipage}%
	\begin{minipage}{.5\linewidth}
		\centering
		\subfloat[][T=400]{\label{cdf sez 400}\includegraphics[scale=.38]{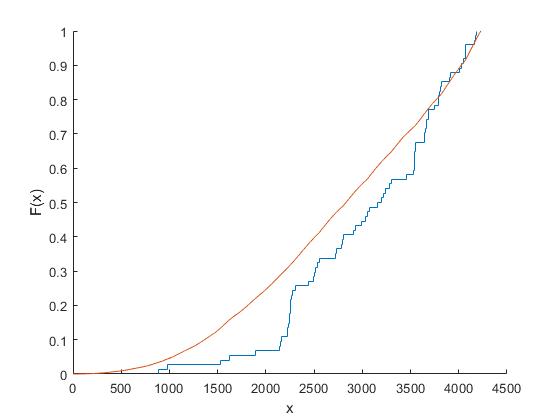}}
	\end{minipage}\par\medskip
	\centering
	\subfloat[][T=700]{\label{cdf sez 700}\includegraphics[scale=.38]{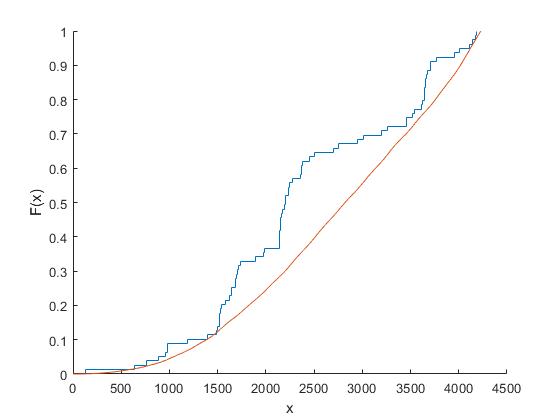}}
	\caption{Comparison between the cumulative distribution function of $F$ and the empirical cumulative distribution function of the jumps arrival times}
	\label{cdf ecdf comparison}
\end{figure}

In order to obtain the p-value in the branching case, we apply the asymptotic results in Facchinetti \cite{Facchinetti2009} in the case when the dataset is greater than $35$.
We provide the $p$-values in the following table.

\begin{table}[H]
	\centering
	\begin{tabular}{ |c|c|c|c| }
		\hline
		Values of $T$ & \quad \quad \quad $200$ \quad \quad \quad & \quad \quad \quad $400$ \quad \quad \quad & \quad \quad \quad $700$ \quad \quad \quad \\
		\hline
		\hline
		Poisson & $\sim 0$ & $\sim 0$ & $\sim 0$ \\
		\hline
		Branching & 0.041 & 0.018 & 0.042 \\
		\hline
		Hawkes & 0.23 & 0.31 & 0.13	\\
		\hline
	\end{tabular}
	\caption{p-value test for the three models for different $T$}
	\label{tab: p_value_estimation}
\end{table}

As one can see from Table \ref{tab: p_value_estimation}, the hypothesis that the intensity follows a Poisson process is highly rejected, as we expected. In the branching case the hypothesis is also rejected, even if the $p$-value in this case was much closer to the acceptance level of $0.05$.
For the Hawkes case the test fails to reject the hypothesis since all the three values are above our level of acceptance.

\bigskip

As a conclusion we can resume the main achievement presented in the present paper. We proposed two alternative models for power forward prices evolution, based on a HJM approach, extending to forward prices dynamics two models already proposed for the spot price dynamics \cite{Eyjolfsson2018}, \cite{JMSS} . After extracting forward curves from quoted futures prices, we proposed a parameters estimation method for both models and then we performed a test on the adequacy of the two models in describing the observed forward prices evolution. The final conclusion of our test is that the hypothesis that forward prices follow a CBI-type dynamics is rejected, while the hypothesis of a Hawkes type dynamics is not. This conclusion suggests that self-exciting effects can arise in power forward dynamics as well as in the spot dynamics, and that an approach based on Hawkes processes can capture these effects in a natural and parsimonious way.


\begin{thebibliography}{99}
	
\bibitem{ADA1} Anderson, F. B., and Deacon, M. (1996): Estimating and Interpreting the Yield Curve. \textit{John Wiley and Sons}.

\bibitem{BKO07} Benth, F. E., Koekebakker, S. and Ollmar, F. (2007): Extracting and Applying Smooth Forward Curves From Average-Based Commodity Contracts with Seasonal Variation, \textit{Journal of Derivatives}, 15(1), 52-66.

\bibitem{BP17} Benth, F. E., Paraschiv, F.  (2018): A space-time random field model for electricity forward prices, \textit{Journal of Banking and Finance}, 95, 203-216.

\bibitem{BPV18} Benth, F. E., Piccirilli, M. and Vargiolu, T. (2019): Mean-reverting additive energy forward curves in a Heath-Jarrow-Morton framework, \textit{Mathematics and Financial Economics}, 13(4), 543-577.

\bibitem{BBK} Benth, F. E., Salthyte-Benth J. and Koekebakker S.  (2008): \textit{Stochastic Modelling of Electricity and Related Markets }, World Scientific, Singapore.

\bibitem{BerSalSco} Bernis, G., Salhi, K. and Scotti, S. (2018): Sensitivity analysis for marked Hawkes processes: application to CLO pricing, \textit{Mathematics and Financial Economics}, 12(4), 541–559.

\bibitem{Bernis_Scotti_Sgarra} Bernis, G., Scotti, S. and Sgarra, C. (2019): A Gamma Ornstein-Uhlenbeck model driven by a Hawkes process, preprint, available at SSRN.

\bibitem{CHL2009} Christensen, T.M., Hurn A.S. and Lindsay, K.A. (2009): It never rains, but it pours: modelling the persistence of spikes in electricity markets,  \textit{Energy Journal} 30(1), 25-48.

\bibitem{CFH2013} Clements, A., Fuller J. and Hurn, A.S. (2013): Semi-parametric forecasting of spikes in electricity prices,  \textit{Economic Records} 89(287), 508-521.

\bibitem{DL06} Dawson, D.A. and Li, Z. (2006): Skew convolution semigroups and affine Markov processes. \textit{Annals of Probability}, 34(3), 1103-1142.

\bibitem{DawsonLi} Dawson, A and Li, Z. (2012): Stochastic equations, flows and measure-valued processes. \textit{Annals of Probability}, 40(2), 813-857.

\bibitem{DFS2003} Duffie, D., Filipovi\'{c}, D. and Schachermayer, W. (2003): Affine processes and applications in finance, \textit{Annals of Applied Probability}, 13(3), 984-1053.

\bibitem{ErraisGieseckeGoldberg} Errais, E.,  Giesecke, K. and Goldberg, L.R. (2010): Affine Point Processes and Portfolio Credit Risk, SIAM Journal on Financial Mathematics, 1(1), 642-665.

\bibitem{Eyjolfsson2018} Eyjolfsson, H., Tj{\o}shteim, D. (2018): Self-exciting jump processes with applications to energy markets, Ann. Inst. Stat. Math., 70(2), 373-393.

\bibitem{Facchinetti2009} Facchinetti, S. (2009): A procedure to find exact critical values of Kolmogorov-Smirnov test, Italian Journal of Applied Statistics, 21(3-4), 337-359.

\bibitem{Filimonov} Filimonov, V., Bicchetti, D., Maystre and N., Sornette, D. (2014): Quantification of the high level of endogeneity and structural regime shifts in commodity markets, Journal of International Money and Finance, 42(C), 174-192.

\bibitem{F01} Filipovi\'{c}, D. (2001): A general characterization of one factor affine term structure models, Finance and Stochastics, 5(3), 389-412.

\bibitem{Fleten03} Fleten, S.E. and Lemming J. (2003): Constructing forward price curves in electricity markets, Energy Economics, 25(5), 409–424.

\bibitem{FL10} Fu, Z. and Li, Z. (2010): Stochastic equations of non-negative processes with jumps, Stochastic Processes and their Applications, 120(3), 306-330.

\bibitem{HainautMoraux2019} D. Hainaut and F. Moraux (2019): A switching self-exciting jump diffusion process for stock prices, Annals of Finance, 15(2), 267-306.

\bibitem{Hawkes} Hawkes, A. G. (1971): Spectra of Some Self-Exciting and Mutually Exciting Point Processes, Biometrika, 58(1), 83-90.

\bibitem{HL15} He, X. and  Li, Z. (2015): Distributions of jumps in a continuous-state branching process with immigration, Journal of Applied Probability, 53(4), 1166-1177.

\bibitem{HJM} Heath, D., Jarrow, R. and  Morton, A. (1992): Bond Pricing and the Term Structure of Interest Rates: a New Methodology for contingent Claim Valuation, Econometrica, 60(1), 77-105.

\bibitem{HerGon2014} Herrera, R. and  Gonzalez, N. (2014): The modeling and forecasting of extreme events in electricity spot markets, International Journal of Forecasting, 30(3), 477-490.

\bibitem{JMSS} Jiao, Y., Ma, C., Scotti, S., Sgarra, C. (2016): A Branching Process Approach to Power Markets, Energy Economics, 79, 144-156.

\bibitem{KaW71} Kawazu, K. and Watanabe, S. (1971): Branching processes with immigration and related limit theorems, Theory Probab. Appl., 16(1), 36-54.

\bibitem{Kiesel} Kiesel, R. and Paraschiv, F. (2017): Econometric Analysis of 15-minutes intraday electricity prices, Energy Economics, 64, 77-90.

\bibitem{KieselParaschiv2019} Kiesel, R., Paraschiv, F., S{\ae}ther{\o}, A. (2019): On the construction of hourly price forward curves for electricity prices, Computational Management Science, 16(1-2), 345-369.

\bibitem{LC16} Lallouache, M. and Challet, D. (2016):  The limits of statistical significance of Hawkes processes fitted to financial data, Quantitative Finance, 16(1), 1-11.

\bibitem{LPV18} Latini, L., Piccirilli, M. and Vargiolu, T. (2018):  Mean-reverting no-arbitrage additive models for forward curves in energy markets, Energy Economics, 79, 157-170.

\bibitem{Li11} Li, Z. (2011): \textit{Measure-Valued Branching Markov Processes}. Springer, Berlin.

\bibitem{LM13} Li, Z. and Ma, C. (2015):  Asymptotic properties of estimators in a stable Cox-Ingersoll-Ross model, Stochastic Processes and their Applications, 125(8), 3196-3233.

\bibitem{Ozaki1979} Ozaki, T. (1979): Maximum Likelihood Estimation of Hawkes self-exciting point processes, Ann. Inst. Statist. Math., 31(1), 145-155.

\bibitem{Paraschiv2013} Paraschiv, F. (2013): Price dynamics in electricity markets, in Handbook of Risk Management in Energy Production and Trading, pp. 47-69.

\bibitem{Pardoux2016} Pardoux, E. (2016): \textit{Probabilistic Models of Population Evolution}, Springer, Berlin.

\bibitem{Rambaldi2015} Rambaldi, Q., Pennesi, X. and Lillo, F. (2015): Modeling foreign exchange market activity around macroeconomic news: Hawkes process approach, Phys. Rev. E, 91(1), 012819.

\bibitem{Schwarz}  Schwarz, E.S. (1997): The stochastic behaviour of commodity prices: implications for valuation and hedging, Journal of Finance, 52(3), 923-973.

\bibitem{Walsh1980} Walsh, J. (1980): \textit{An Introduction to Stochastic Partial Differential Equations}.  Ecole d'\'{e}t\'{e} de Probabilit\'{e}s de Saint-Flour XIV-1984, Lecture Notes in Mathematics 1180, 265-430. Springer, Berlin.




\end{thebibliography}
\end{document}